\documentclass[reprint, amsmath,amssymb,
 aps,nofootinbib,showkeys, twocolumn]{aastex63}
\usepackage{color}

\usepackage{amsmath}

\submitjournal{ApJ}

\shorttitle{Consequences for the SFDM model from the McGaugh Observed-Baryon Acceleration Correlation}
\shortauthors{Padilla et al.}

\begin{document}

\title{Consequences for the Scalar Field Dark Matter model from the McGaugh Observed-Baryon Acceleration Correlation}

\correspondingauthor{Luis E. Padilla}
\email{epadilla@fis.cinvestav.mx}

\author[0000-0002-5681-7699]{Luis E. Padilla}
\affiliation{Instituto de Ciencias F\'isicas, Universidad Nacional Aut\'onoma de M\'exico,
Apdo. Postal 48-3, 62251 Cuernavaca, Morelos, M\'exico.}
\affiliation{Mesoamerican Centre for Theoretical Physics,
Universidad Aut\'onoma de Chiapas, Carretera Zapata Km. 4, Real
del Bosque (Ter\'an), Tuxtla Guti\'errez 29040, Chiapas, M\'exico}

\author[0000-0001-5564-7133]{Jordi Sol\'is-L\'opez}
\affiliation{Departamento de F\'{i}sica, Centro de Investigaci\'on y de Estudios Avanzados del IPN, A.P. 14-740, 07000 CDMX, M\'exico.}
\author[0000-0002-0570-7246]{Tonatiuh Matos}
\affiliation{Departamento de F\'{i}sica, Centro de Investigaci\'on y de Estudios Avanzados del IPN, A.P. 14-740, 07000 CDMX, M\'exico.}
\author[0000-0003-2223-4716]{Ana A. Avilez-L\'opez}
\affiliation{Facultad de Ciencias F\'isico-Matem\'aticas, Ciudad Universitaria, Benem\'erita Universidad Aut\'onoma de Puebla, Av. San Claudio SN, Col. San Manuel, Puebla, M\'exico. }

\begin{abstract}

Although the standard cosmological model, the so-called $\Lambda$ cold dark matter ($\Lambda$CDM), appears to fit well observations at the cosmological level, it is well known that it possesses several inconsistencies at the galactic scales. In order to address the problems of the $\Lambda$CDM on a small scale, some alternative models have been proposed. Among the most popular candidates, the proposal that dark matter in the universe is made of ultralight bosons is a strong candidate today.
For this work, we study through an analytical approach the consequences arising from comparing the SPARC catalog observed-baryon acceleration correlation with the scalar field dark matter model. We carry out such analysis either considering the features of galactic halos extracted from structure formation simulations or considering the existence of other non-dark-matter elements in the whole system (such as baryons or a supermassive black hole). 
Specifically, we address a recent claim that the model is not capable of reproducing a constant surface density in the core, in contrast to what observations suggest for a host of galaxies with different sizes and morphologies. In this direction, we show that this discrepancy can be alleviated once the contributions of non-dark-matter constituents in the whole galactic system are taken into account. Additionally, we find that a mass of $m\simeq 1.41\times 10
^{-22}\ \rm{eV}/c^2$ is capable of reproducing all of our findings and correctly adjusting the rotation curves coming from the Milky Way galaxy.

\end{abstract}


\section{Introduction} 

{In recent years, observations of rotation curves of a large sample of galaxies of diverse sizes and morphologies have been performed. One good example is the Spitzer Photometry and Accurate Rotation Curves (SPARC) catalog given in \cite{McGaugh:2016PRL}. From this catalog it has been possible to derive a strong correlation between the observed acceleration $g_{\rm{obs}}$ and the Newtonian acceleration of baryons $g_{\rm{bar}}$. According to this relation, in the large-acceleration regime, laying above {$g^\dagger=1.2\times 10^{-10}\text{m s}^{-2}$}, which typically corresponds to accelerations in central regions of galaxies, $g_{\rm{obs}}$ can be predicted exclusively from the Newtonian acceleration of baryons. This may imply that the assumption of dark matter (DM) is not required. However, in the outer low-acceleration regime of the galaxy, the correlation indicates that either a modification of the Newtonian acceleration or an extra gravitational pull due to DM is needed.} By holding the second hypothesis, the McGaugh correlation importantly restricts not only the radial acceleration profiles but also the spatial distribution of DM particles.

A first study exploring the consequences of this result has been done  in \cite{Urena-Robles-Matos:2017}. In that work, the authors realized that the McGaugh's correlation implies a {link} between the accelerations produced by DM particles and baryons, which additionally, for every DM core model, brings up a universal upper bound for the maximum acceleration produced by DM particles. This realization sets strong restrictions on the value of the central surface density of the halo. More specifically, for any core DM model in which a two-parameter spherical density profile is assumed -- a characteristic length  $r_s$ and a central density value $\rho_s$ -- it is predicted that the product $\rho_s\,r_s \equiv \mu_{\rm{DM}}$ must be {always constant}. This conclusion has been robustly proven in different types of galaxies, such as spirals of late and early type, dwarf irregulars, ellipticals, and the Local Group dwarf spheroidal (dSph) galaxies \citep[see, for example,][]{spano2008ghasp,donato2009constant,scalingmu1,scalingmu2}, suggesting that it should result in being a universal quantity in all kinds of galaxies.

In the same work, {a particular DM model was studied in detail,  which proposes that DM in the universe is made of ultralight boson particles. In that work, the McGaugh results have an important implication for particular models of this scalar field DM (SFDM)}. By fitting the accelerations of satellite dSph galaxies of the Milky Way (MW), along with some scaling relations of the equations used to describe the SFDM and the constancy of $\mu_{\rm{DM}}$, they predicted that all these systems have a total mass {$M_{\rm{DM}}(300\ \text{pc})\sim 10^7 M_\odot$} and a characteristic size  {$r_s = 300\ \text{pc}$}. This  universal central distribution of DM for dSphs was dubbed as a universal SFDM soliton for such systems. Surprisingly, this prediction is in agreement with the results reported in \cite{NatureBullock}. Together the results of \citet{McGaugh:2016PRL} and \citet{Urena-Robles-Matos:2017} could lead to probing the hypothesis that the SFDM soliton, under certain conditions, has universal features where the physics due to extragalactic components does not deform or destroy the soliton. Regarding to this point, it is worth mentioning that, within the SFDM model, some scenarios have been considered where more general initial conditions are taken into account, and consequently some dynamical features arise  that prevent the soliton formation within times shorter than the age of the universe, by gravitational cooling. Therefore, testing the implications of the existence of the soliton also brings up valuable information for testing the dynamical aspects of this sort of DM \citep{Avilez-Lopez:2018hwh,sc14}.
It is clear that this hypothesis must undergo through further observational probes in the future, but, at this date, it stands as an interesting remark regarding the SFDM model. 

It is worth dedicating some words to stressing that {the SFDM model is a strong and well-studied alternative to the {standard cold dark matter} (CDM) paradigm. The main idea that scalar fields {might make up dark matter} in the universe was originated about two decades ago \citep{sf5,sf1,sf2,sf3,sf40,sf4,sf6,sf7}, although some hints can be traced further back in \cite{sf80} and \cite{sf8}.  Since then, the idea has been rediscovered by various authors with different names, for example: SFDM \citep{sf4}, fuzzy DM \citep{sf2}, wave DM \citep{sc4}, Bose-Einstein condensate DM \citep{sf10} or ultralight axion DM \citep{sf11} (see also \cite{sf12}). However, it was systematically studied for the first time by \cite{sf13} and \cite{sf14};  \citep[for a review of SFDM, see][]{rev1,rev2,rev3,rev4,niemeyer2019small,RS}. In this work, we use the most usual and general name SFDM.}

{For the scalar field to behave as dust at late times, a quadratic term of the scalar needs to appear} in the Lagrangian: %
\begin{equation}\label{potential_eq1}
    V(\varphi) = \frac{1}{2}\frac{m^2c^2}{\hbar^2}\varphi^2,
\end{equation}
where $m$ is the mass of the bosons, $\hbar$ is the Planck constant, and $c$ is the speed of light. This potential term gives rise to a pressureless fluid behavior during the matter-dominated epoch, {where SFDM is the dominant component.} The astrophysical motivation for this model has its root in the small-scale 
problems of CDM. This model alleviates such problems as thanks to its dynamical properties derived from the macroscopic-sized de Broglie wavelength associated with its bosonic components. In a host of different {cosmological simulations of structure formation} \citep{sc4,sc10,sc11,sc12,sc13,sc14} it has been shown that SFDM halos have cored density profiles within their inner regions {of galactic systems;} this feature naturally solves the cusp/core problem in CDM. {These cores, referred to as ``solitons" in the literature \citep{sc14,sc15,sc16,sc17} have been shown to have a size that is {similar in magnitude to} the de Broglie wavelength of individual bosons:
\begin{equation}\label{xdb}
    \lambda_{dB}\propto \frac{1}{mv},\nonumber
\end{equation}
where $v$ is the ``{average} virial velocity" of the bosons,  a result expected from analytic calculations. To reproduce galactic cores of order 1 kpc for the model, the boson mass is typically assumed to be in the range of $m\sim (10^{-20}-10^{-22})~\rm{eV}/c^2$. However, 
these cores have been found to be surrounded by {an envelope generated by quantum interference patterns inherent to SFDM that is well fitted by a Navarro-Frenk-White (NFW) density profile}. From simulations made by \cite{sc4,sc10}, it was realized that the masses of the core and the whole halo {obey a strong scaling correlation given by} 
\begin{equation}\label{mcmh}
M_c\propto M_h^{1/3}, \nonumber
\end{equation}
where $M_c$ and $M_h$ are the total core and halo mass, respectively. Then, the minimum mass of galaxies that can be generated in this model {satisfies} $M_c = M_h$ solving naturally the small-scale problem of CDM. This {scaling relation between the masses of the core} and its ``envelope" has not been anticipated by any earlier work, though it is not difficult to understand the form of the correlation in an \textit{a posteriori} way, using analytic arguments. Indeed, the fact that this correlation has been robustly established by simulations} offers a unique opportunity to understand and to extend the correlation by considering novel physical situations, such as the addition of {further} matter contributions in the total halo.

One of the problems {for this model} that is currently becoming more and more evident is the fact that {the results} from cosmological simulations of galaxy formation do not seem to agree with the results of a constant surface density, as suggested by \cite{spano2008ghasp,donato2009constant,scalingmu1} and \cite{scalingmu2}. Thus, it has been claimed that the SFDM is not able {to provide a correct mechanism to describe the core-formation mechanism in galaxies in a realistic way} \citep[see, for example,][]{nomu1,nomu2}. However, most of the works regarding this model and the conclusions mentioned here until now have been commonly obtained from DM-only simulations. {By keeping this in mind, a second goal in this work is to extend such DM-only results in order to take into account the effect of further matter contributions besides to DM in the system within an analytical scheme. Considering such extra components is} an important step in modeling galactic systems.

{For instance, a recent attempt to study how a central supermassive black hole (SMBH) may deform the SFDM density profile can be found in \cite{Avilez-Lopez:2017zfp}.}
Usually, radial DM and baryonic density profiles are used independently when deriving the stellar kinematics; however, since both are shaped by the total gravitational potential, it would be more accurate to have a connected pair of profiles for baryons and DM rather than modeling them separately. In that line, a goal of this work is to further explore the implications of the McGaugh $g_{\rm{obs}}(g_{\rm{bar}})$ correlation over the density profiles of SFDM {in systems where more matter contributions are present.} Our goal is to demonstrate, by considering some analytical relations, that once after the mentioned extra components are considered, the tension between the results from SFDM simulations and the result of constant surface density is alleviated. Furthermore, we find out that the scaling core-halo mass relations hold in this more general scenario. Afterward, we show that SFDM is able to agree with observations.

The paper is organized as follows. In section \ref{section2} we present some general implications for the McGaugh observed-baryon acceleration correlation. Particularly, we show that such a correlation implies a maximum possible acceleration produced by any DM candidate inside a galactic halo. Using this result, {we probe a host of DM profiles that are usually derived from DM-only simulations.} Then, we discuss in a more specific way the consequences that are obtained for our SFDM model. In section \ref{section3} we begin by detailing the system of equations that are used to describe galaxies {within the SFDM model to later study the possibility of extending the mass-ratio relation of the soliton solution after further  matter contributions} are added to the system. For such purpose, we adopted a Gaussian ansatz to describe {the density profile of the soliton}. Later in this section, we explore the possibility of extending the core-halo mass relation {derived from numerical SFDM-only simulations.} In section \ref{section4} we compare our previous findings with the McGaugh correlation function, and then we extend the results presented in section \ref{section2}. In the same section, we test all of the conclusions that we reached in our work by comparing them with observational data from the Milky Way. We find that a mass of $m\sim 10^{-22} \rm{eV}/c^2$ is able to fit all of our results and correctly describe the observations. Finally, in section \ref{section5} we present our conclusions.

\section{Observational Correlation of Baryonic and Dark Matter {Accelerations} Observational Correlation }\label{section2}

\subsection{General implications}\label{sectionII}

In this section, we review some theoretical implications of the observational correlation between the accelerations of DM and baryonic components found by \citet{McGaugh:2016PRL}.  {This phenomenological relation implies some conditions} for some features of the DM spatial distribution and its kinematics. In \cite{McGaugh:2016PRL}, by analyzing the high-precision data from 153 spiral galaxies in the SPARC database, {the authors} found a correlation between the acceleration observed in stars, $g_{\rm{obs}}$, and that inferred to be produced by baryons, $g_{\rm{bar}}$. This empirical correlation, usually known as the radial acceleration relation (RAR), is given by 
\begin{eqnarray}
g_{\rm{obs}}=\frac{g_{\rm{bar}}}{1-e^{-\sqrt{g_{\rm{bar}}/g^\dagger}}}.
\label{eq:gMcGaugh}
\end{eqnarray}
\citet{lelli2017one} have further established that a similar relation holds for other types of galaxies such as ellipticals, lenticulars, and dSphs, suggesting that the above RAR could be thought of as a universal correlation {fulfilled by} different types of galaxies, and then {it provides} an unprecedented test for DM models. In order to draw this conclusion more {clearly}, {it is interesting to notice} that, depending on the value of the acceleration observed in stars $g_{\rm{obs}}$, {that produced by baryons $g_{\rm{bar}}$ suffices to reproduce such observations in a well-defined acceleration regime.} In the case $g_{\rm{obs}}>g^{\dagger}\equiv 1.2\times 10^{-10} \rm{ms^{-2}}$, the acceleration produced by baryons can describe correctly that observed in stars, whereas in the opposite case $g_{\rm{obs}}<g^\dagger$, {an extra hypothesis needs to be considered} to explain the observations. This suggests that $g^\dagger$  serves to discriminate between {two regimes where the acceleration of matter inside the galaxy is mainly produced only by baryons and gravity is described by general relativity or not.} The previous argument can be translated into realistic scenarios. For example, as pointed out in \cite{lee2019radial}, there are three regions {within the sampled galaxies} where $g_{\rm{obs}}$ can be much smaller than $g^\dagger$: (1) outermost edge of galaxies ($r> O(10^2)\ \rm{kpc}$), (2) the outer parts of the disc of massive galaxies where stars follow flat rotation curves ($\rm{kpc}<r< O(10)\ \rm{kpc}$), and (3) small dwarf galaxies ($r<\rm{kpc}$). In the opposite case, in massive galaxies it is usually expected that in their central regions happens that $g_{\rm{obs}}\gg g^\dagger$ and then it is largely dominated by baryons.

In a more recent work \citep{salucci}, it has been argued that the relation (\ref{eq:gMcGaugh}) is not {completely}  general. {In that work, it is shown} that dwarf disc spirals and low-surface-brightness galaxies do not follow the same $g_{\rm{obs}}-g_{\rm{bar}}$ relation as other types of galaxies, suggesting that McGaugh's et al. results would actually turn out to be a limiting case of a more general relationship. {In this work, the results are derived assuming the McGaugh observed-baryon acceleration correlation, so they are only applicable for those galaxies that fulfill with Equation (\ref{eq:gMcGaugh}).}

\begin{figure}[t!]
\centering
\includegraphics[width=3.4in]{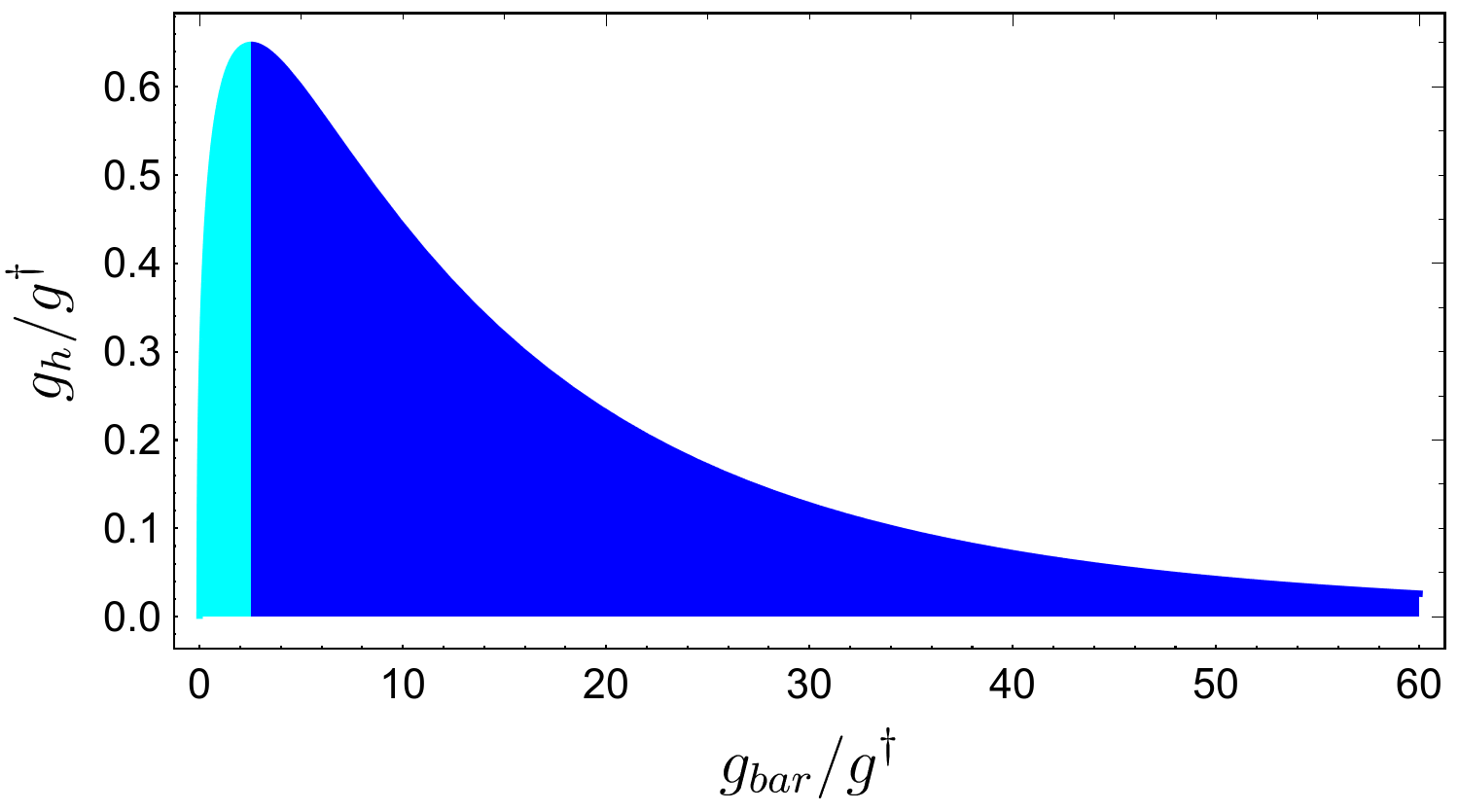}
\caption{Inferred correlation between accelerations produced either by DM and by baryonic particles in units of $g^\dagger$. Dark blue region correspond to regions close to the galactic center at $r\lessapprox 8$ kpc. Acceleration of DM particles in this region decrease implying that at the very center they tend to behave as free particles.}
\label{fig:ghgb}
\end{figure}

If we take $g_{\rm{obs}} = g_{\rm{bar}}+g_h$, where $g_h$ is the radial acceleration produced by nonbaryonic matter, Equation (\ref{eq:gMcGaugh}) leads  straightforwardly to an even more interesting correlation between $g_h$ and $g_{\rm{bar}}$, given by
\begin{eqnarray}\label{g_h_max}
g_h&=&\frac{g_{\rm{bar}}}{e^{\sqrt{g_{\rm{bar}}/g^{\dagger}}}-1}.
\label{eq:gh}
\end{eqnarray}
The above expression has a maximum given by  
\begin{equation}\label{g_maximum}
g_{h,max} = 0.65 g^\dagger\ \ \text{at}\ \  g_{\rm{bar}}=2.54\, g^\dagger,
\end{equation}
(see figure \ref{fig:ghgb}). This is interesting because it is well known that the main contribution to the radial acceleration in a galaxy (apart from baryons) is produced by a DM component $g_{\rm{DM}}$ and possibly by a SMBH at galactic nuclei $g_{\rm{SMBH}}$\footnote{A SMBH is expected to exist in almost all galaxies, except for small dSphs.}, that is, $g_h = g_{\rm{DM}}+g_{\rm{SMBH}}$, where the contribution of the last one is only important within the innermost regions of galaxies. The fact of the existence of this maximum  acceleration $g_{h}$ and that the acceleration produced by the central SMBH increases enormously for small radius ($g_{\rm{SMBH}} = GM_{\rm{BH}}/r
^2$, where $G$ is the Newton's constant, $M_{\rm{BH}}$ is the mass of the SMBH, and $r$ is the radial coordinate measured from the position of the SMBH) suggest that the McGaugh scaling relation does not apply in regions along the galaxy where the effect of the SMBH dominates over all of the other components. This assumption implies that $g_h = g_{\rm{DM}}$, and then $g_{h,max}$ {should result in a maximum possible acceleration produced by any DM model}. Such a conclusion will be especially helpful here to probe the predictions of the SFDM according to the McGaugh results.

Note that Equation (\ref{g_h_max}) could serve as a tracer of the spatial distribution for DM along the galaxy. That is, if we know the functional form $g_{\rm{bar}}$ for baryons, from Equation (\ref{g_h_max}) we could infer the density of DM particles $\rho_h$ as
\begin{equation}
\rho_{h}=\frac{1}{4 {\pi} G}\frac{1}{r^2}\frac{d}{dr}(r^2g_h(g_{\rm{bar}}(r))).
\end{equation}
An interesting consequence from the last expression is that the McGaugh correlation leads to a model-independent DM density profile once a baryonic distribution in a galaxy is known. Particularly, observe the fact that the acceleration of DM is not an increasing function of $g_{\rm{bar}}$, and hence $g_{h}$ decreases down to the center, which implies that DM particles in such a central region are influenced by a decreasing total force; that is, DM particles in the center tend to behave as free  particles.

\subsection{Implications for DM-only simulations}\label{implicationB}

Although in this work our goal is to explore the implications of the McGaugh observational relation for the SFDM model once the contribution of baryonic matter is taken into account, {in this subsection} we {aim to } probe results coming from DM-only simulations. Clearly,  the best candidates for which this constriction applies are typically dSphs because their mass-to-light ratios suggest that they are DM-dominated systems at all radii\footnote{Such an inference is compatible with  the McGaugh et al. results since for such systems it happens that $g_{\rm{obs}}\ll g^{\dagger}$}. {Additionally, \citet{lelli2017one} have shown that these kinds of galaxies (at least those belonging to the Local Group) also fulfill Equation (\ref{eq:gMcGaugh})}. In this subsection, we review and complement the study reported in  \cite{Urena-Robles-Matos:2017} because it will be very instructive for understanding the next sections, where we reach conclusions additional to those obtained by Ure\~na et al.

As realized by  \citet{Urena-Robles-Matos:2017}, once a spherical density profile for DM $\rho_{\rm{DM}}$ is assumed, it can be expressed, without lost of generality, as $\rho_{\rm{DM}}(r) = \rho_s f(r/r_s)$, where $\rho_s$ and $r_s$ are the characteristic density and radius of the DM profile, respectively, and $f(r/r_s)$ is a dimensionless function characterizing the spatial distribution of DM. For the particular case of cored-like profiles, $\rho_s$ is the central value of the density, and therefore $f(0) = 1$. Thus, the acceleration produced by DM, {$g_{\rm{DM}}(r)= GM(r)/r^2$, where $M(r)= \int_0^r 4\pi \rho_{\rm{DM}}(r')r'^2 dr'$ is the DM mass at a distance $r$, can be written as}
\begin{equation}\label{g_dm}
    g_{\rm{DM}}(r) = G\mu_{\rm{DM}}\hat g_{\rm{DM}}(x),
\end{equation} 
{where we have introduced the dimensionless acceleration function}
\begin{equation}
\hat g_{\rm{DM}}(x) \equiv\frac{4\pi}{x^2} \int_0^x f(x') x'^2 dx',
\end{equation}
with $x\equiv r/r_s$, and 
\begin{equation}\label{surface_density_definition}
\mu_{\rm{DM}}\equiv \rho_s r_s.  
\end{equation}
Notice that Equations (\ref{g_maximum}) and (\ref{g_dm}) imply the following condition for the maximum acceleration:
\begin{equation}\label{mu_const}
    \frac{0.65 g^\dagger}{10^{-11}\rm{ms^{-2}}} = 0.014\left(\frac{\mu_{\rm{DM}}}{M_\odot \rm{pc^{-2}}}\right)\hat g_{DM,max}.
\end{equation}
This equation implies that the surface density parameter remains constant within any DM model as $\hat g_{DM,max}$ also remains constant. Such a realization holds within any DM model and for any type of galaxy {that fulfills Equation (\ref{eq:gMcGaugh})}. Furthermore, it is an inference that has been previously suggested by a host of observations \citep[see, for example,][]{spano2008ghasp,donato2009constant,scalingmu1,scalingmu2}. Consequently, for a particular DM density profile, once the dimensionless maximal acceleration $\hat g_{h,max}$ is computed, it is straightforward to constrain the surface density parameter $\mu_{\rm{DM}}$ within such model. It is worth stressing that, until now, this prescription is fully general and, as mentioned above, it can be applicable to describing all kinds of galaxies {that fulfill Equation (\ref{eq:gMcGaugh})}. 

\begin{figure}
\includegraphics[width=0.4\textwidth]{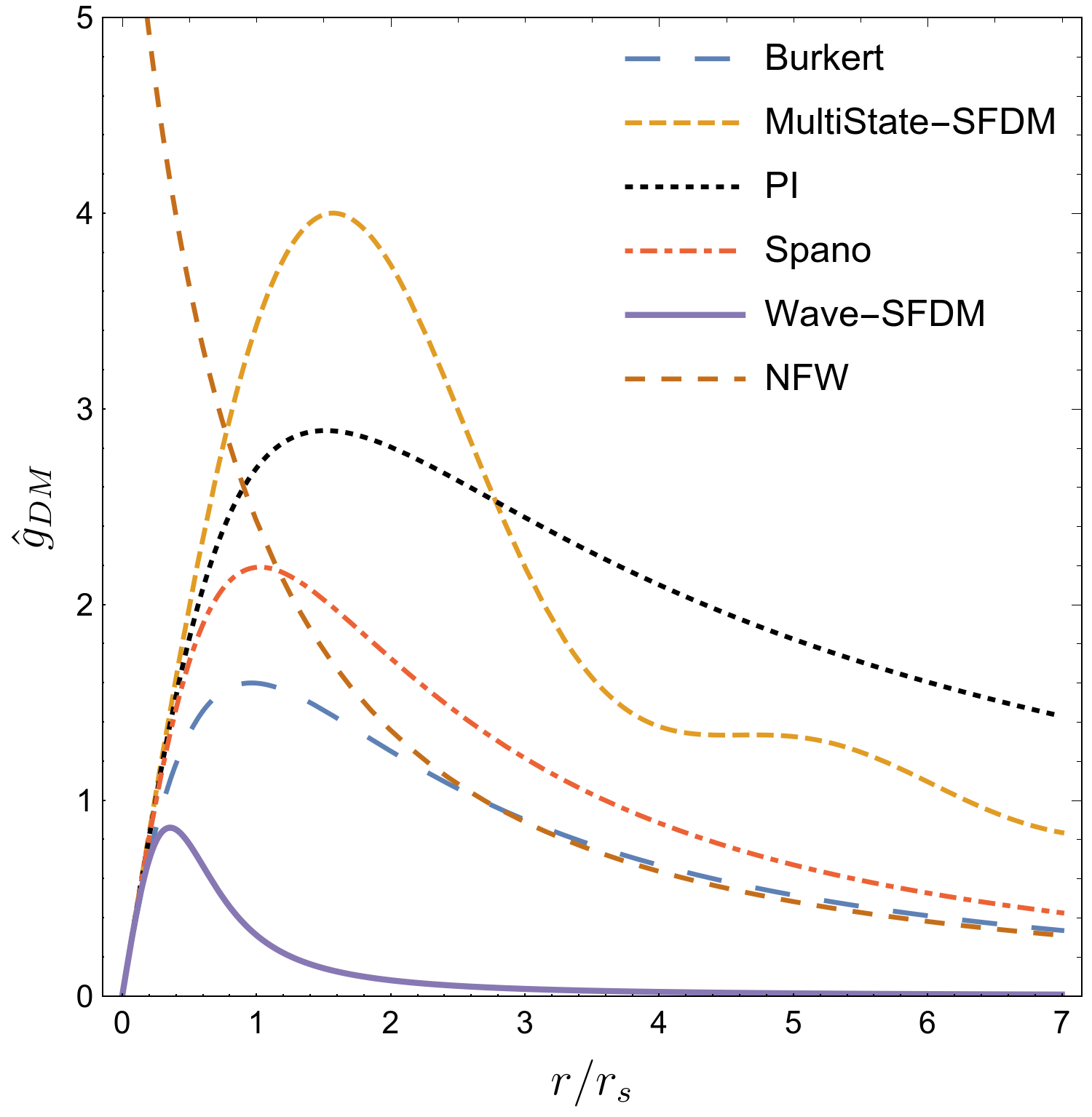}
\caption{The dimensionless acceleration profiles $\hat g_{\rm{DM}}$ corresponding to the Burkert, MultiState-SFDM, pseudoisothermal (PI), Spano, wave-SFDM and NFW  density profiles as functions of the dimensionless radius $r/r_s$.}
\label{fig:Aceleras}
\end{figure}

Now, let us analyze Equation (\ref{mu_const}) more closely by considering some particular DM profiles. For instance, we take the Burkert \citep{Burkert:1995}, the pseudoisothermal (PI) \citep{PI:1991}, the Spano \citep{spano2008ghasp}, the NFW, and two well-known SFDM density phenomenological profiles that are valid for describing dSphs. For the SFDM instance, we consider firstly the so-called wave-SFDM density profile \citep{sc4,sc10}, which is described as radial time-averaged 3D solutions of the Schr\"odinger-Poisson (SP) system. Second, we consider the MultiState-SFDM profile derived  analytically in \cite{robles2012exact} by solving the Klein-Gordon equation in a Minkowski background space-time. For all these models, either the dimensionless density $f(x)$ or the dimensionless acceleration $\hat g_{\rm{DM}}(x)$ is listed in Table \ref{tab:profiles} and plotted in figure \ref{fig:Aceleras}. Interestingly, the acceleration {corresponding to} all cored-like density distributions {starts} from zero at small $x$ and increases up to the maximum acceleration value $g_{DM,max}$ (fourth column in Table \ref{tab:profiles}) at $x_{max}$ (fifth column). With such expressions and Equation (\ref{mu_const}), it is straightforward to compute the surface density $\mu_{\rm{DM}}$ corresponding to each DM model (sixth column). 

\begin{deluxetable*}{cccccccc}
\tablecaption{Density and acceleration profiles for different models of DM. \label{tab:profiles}}
\tablewidth{0pt}
\tablehead{\colhead{Name} & \colhead{$f(x)$} &
\colhead{$ \hat{g}_{\rm{DM}}(x)$}& \colhead{ $ \hat{g}_{DM,max}$} & \colhead{$x_{max}$} & \colhead{$\mu_{\rm{DM}}[M_\odot\rm{pc^{-2}}]$} & \colhead{$r_s[\rm{pc}]$} & \colhead{$\rho_s[M_\odot\rm{pc^{-3}}]$}}
\decimalcolnumbers
\startdata
Burkert & $\frac{1}{(1+x)(1+x^2)}$ & $\frac{\pi \left(2\ln(1+x)+\ln(1+x^2)-2\arctan x\right)}{x^2}$ & 1.59 & 0.96 & 348 & 48;1400 & 3.82;0.23  \\
MultiState-SFDM   & $\frac{\sin^2(x)}{x^2}$ & $\frac{2\pi}{x}\left(1-\frac{\sin(2x)}{2x}\right)$ & 4 & $ \frac{\pi}{2}$ & 139 & 79;851 & 1.76;0.163  \\
PI     & $\frac{1}{1+x^2}$ & $\frac{4\pi}{x^2} \left(x-\arctan x \right)$ & 2.89 & 1.515 & 193 & 29; 11.667 & 6.69; 0.16 \\ 
wave-SFDM   & $\frac{1}{(1+x^2)^8}$ & $\frac{\pi}{x^2} \left(\frac{33}{512}\arctan x+\frac{1}{53760}\frac{P_{\rm{wave}}}{(x^2+1)^7}\right)$ & 0.86 & 0.36 & 648 & 299;3961 & 2.16; 0.163 \\
Spano  & $ \frac{1}{\left(1+x^2\right)^{3/2}}$ & $\frac{4 \pi}{x^2}  \left(\text{arcsinh} x-\frac{x}{\sqrt{x^2+1}}\right)$ & 2.19 & 1.028 & 254 & 68; 1561 & 3.72; 0.163 \\
NFW & $\frac{1}{x(1+x)^2}$ & $\frac{4\pi}{x^2} \left( \ln(1+x)-\frac{x}{1+x}\right)$ & 0 & 6.28 & 89 & / & /  \\
\enddata
\tablecomments{Second column: dimensionless density radial profile. Third column: dimensionless acceleration profile due to DM. Fourth and fifth columns: maximum acceleration value and the radius where such value occurs, respectively. Sixth column: the constant $\mu_{\rm{DM}}=r_s\rho_s$ parameter. Seventh and eighth columns: density profile parameters that reproduce observations. We have defined $P_{\rm{wave}}=3465 x^{13}+23100x^{11} + 65373 x^9 + 101376 x^7 +92323 x^5 + 48580 x^3 -3465x$.}
\end{deluxetable*}

\begin{figure}
    \centering
    \includegraphics[width=0.45\textwidth]{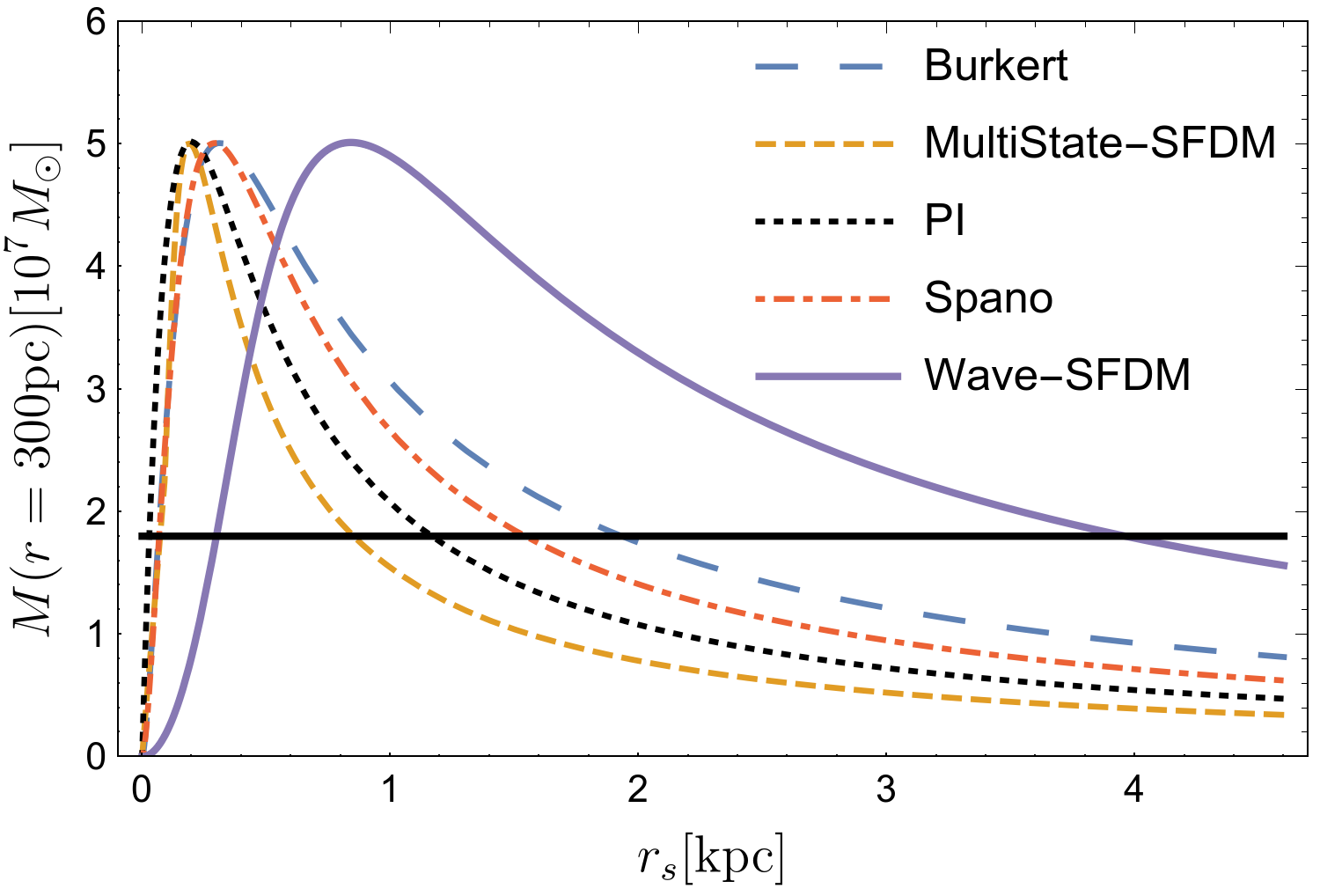}
    \caption{Total mass enclosed within radius $r = 300\ \rm{pc}$ as a function of $r_s$ for different DM models. The horizontal line corresponds to the mean value inferred from \citet{NatureBullock} given by $M(r\simeq300\ \rm{pc})\simeq 1.8\times 10^7\ M_\odot$.}
    \label{fig:mass_300}
\end{figure}

On the other hand, {note that the result of obtaining a constant density in these models allows us to write $r_s = r_s(\rho_s)$ or $\rho_s = \rho_s(r_s)$, and then, if somehow we were able to independently fix any of these two quantities, this would immediately result in being able to fix the value of the other parameter. To get an idea of how to do this, let us consider the following possibility.} In \cite{NatureBullock}, it was suggested that the mass {enclosed} inside $r = 300\ \rm{pc}$ in dSph satellites of the Milky Way {is around }$M(r \simeq 300\ \rm{pc}) \simeq  1.8\ \times 10^7\ M_\odot$\footnote{We obtained this value by averaging over all of the values reported in \cite{NatureBullock}.}. {This was a model-dependent result, {since to obtain this result the authors considered a functional form to describe the dark matter halo. However, due to the generality of the profile they took, \footnote{{The authors modeled the dark matter distribution in galaxies by using the profile
\begin{equation}
    \rho(r) = \frac{\rho_0}{(r/r_0)^a[1+(r/r_0)^b]^{(c-a)/b}},
\end{equation}
where the asymptotic inner slope is determined by $a$, the asymptotic outer slope by $c$, and the transition between these two regimes is determined by $b$.}}, the authors claim that their results are general, however the ranges of the parameters they consider do not cover the wave-SFDM profile. For simplicity and as an example only}, in order to get an idea of the free parameters of each model {and show how we can fix both parameters for each model with only one constriction}, we could assume that this result is true for any model of DM}. {As we already mentioned, }for a given model, $r_s$ and $\rho_s$ are dependent parameters since  $\mu_{\rm{DM}}$ must take the constant value shown in Table \ref{tab:profiles}, and then only one of them is to be fixed in order to compute the total mass enclosed within $r=300\ \rm{pc}$ (see figure \ref{fig:mass_300}). Note that the mass enclosed at such radius coincides with the observations of \citet{NatureBullock} for two different values of $r_s$, which are listed in Table \ref{tab:profiles} for each case. Particularly, for the wave-SFDM model (or only SFDM model from now on) the characteristic core sizes predicted for such galaxies are $r_s = 299\ \rm pc$ and $r_s = 3.961\ \rm{kpc}$. {Of course, these core size values come from this simple example that we did, and they should not necessarily be correct for the model, since to be sure, a study similar to the one done in \cite{NatureBullock} should be done once adopting the dark matter profile that describes a halo in SFDM.} 

Finally, owing to a scaling symmetry of the SP system, the SFDM model holds a scaling property for $\rho_s$ and $r_s$ that works as follows: $\rho_s = \lambda^4 m^2 m_{pl}^2/4\pi$ and $r_s = (0.23 \lambda m)^{-1}$, where $m_{pl}$ is the reduced Planck mass. By eliminating the scaling parameter $\lambda$, we finally arrive at the expression
\begin{equation}\label{mu_constant}
    \left(\frac{r_s}{\rm{pc}}\right)^{-3}(m_{22})^{-2} = 4.1\times 10^{-13}\left(\frac{\mu_{\rm{DM}}}{M_\odot\rm{pc^{-2}}}\right),
\end{equation}
where $m_{22} \equiv m/(10^{-22}\rm{eV}/c^2) $. Given that $\mu_{\rm{DM}}$ has been found to be a constant quantity in this model from the description we have followed until now, this result predicts that $r_s$ should also be a constant quantity and scale as $r_s\propto m^{-2/3}$. {It should be stressed that these scaling relations are only valid for DM-dominated systems such as dSph, and they need to be further studied when external sources such as baryons are included, as we aim to do in this work.}  Additionally, if we use the values in Table \ref{tab:profiles}, we can finally {obtain an estimate of } the mass of the SFDM boson {particle (of course, valid only for our example)}, given by
\begin{equation}\label{m_22_300}
    m_{22} \simeq 0.246 \ \ \ \text{and}\ \ \ m_{22} \simeq 11.8,
\end{equation}
{where the first constraint follows from $r_s =3.961 \ \rm{kpc}$ and the second from $r_s = 299\ \rm{pc}$. However, these mass estimates must be refined once we compare the SFDM model directly with the observational data coming from dSphs of the Milky Way, {as we already mentioned}. For example, this model has been tested in \cite{lora1} by taking the problem of the longevity of the stellar group present in Ursa Minor or the problem in the orbits of the globular clusters in Fornax. In that work, they could constrain the mass parameter of the SFDM to be in the range $0.3\leq m_{22}\leq 1$ in order to be consistent with observations. A similar result was obtained in \cite{lora2}, where they confronted the SFDM model by exploring how the time scale of dissolution of stellar substructures in Sextans dSph contributes to constrain the free parameters of the model, obtaining the result $0.12\leq m_{22}\leq 8$. Using these last two results together, it is found that the most preferred mass parameter for the SFDM should be in the range $0.3\leq m_{22}\leq 1$. 

{Of course, there exist a large number of other constraints exist that have been provided for the SFDM model, and then we will mention only some representative results. The SFDM model has been studied at the perturbative level, showing that the cosmic microwave background (CMB) can be adjusted perfectly as long as the mass of the SFDM particles fulfil with the condition $m_{22}\geq 10^{-2}$ \citep{free_const8}. Explaining the half-light mass in the ultrafaint dwarfs fits the mass to be $m_{22} \sim 3.7 - 5.6$ \citep{free_const4}. The model has also been constrained by observations of the reionization process. In \cite{free_const5}, using $N$-body simulations and demanding an ionized fraction of HI of 50\% by z = 8, $m_{22} > 0.26$ was obtained. Finally, the Lyman-$\alpha$ forest flux power spectrum
demands that the mass parameter fulfills $m_{22} \geq 20 - 30$ \citep{free_const6,free_const7}.}

{It is not difficult to realize that, even though several constrictions are consistent with each other, this does not happen with all of them, being that for now, constrictions coming, for example, from the Lyman-$\alpha$ forest are strongly in tension with what other groups have reported through other kinds of observations. However, the fact that observations coming from Lyman-$\alpha$ do not fit perfectly with the other constrictions is not necessarily a problem, since this may be because there are effects that are not really being taken into account in the description of the model. For example, a very popular candidate to be this ultralight scalar field particle is that of an axion-like particle with a trigonometric potential that, although effectively it can be described as a field that behaves like a free field, it has been shown that once the full trigonometric potential is taken into account, it is possible to alleviate this discrepancy, since for this case, a mass of $m_{22}\sim 1$ could meet observations from Lyman-$\alpha$ as well \citep[see, for example,][]{free_const9}\footnote{{In this work, we are not interested in studying the fundamental nature that would generate this dark matter candidate, since currently there is a great variety of well-founded potentials that at low energy scales would behave effectively as a free field, that is, subjected to a potential of the form \eqref{potential_eq1}.}}. In this way, we consider that the best way to proceed to continue studying the model is to independently test the SFDM in different scenarios and through different observations. In this way, our intention in this work will be to do just this and to test the model through the results that will be presented later, which will be based on the RAR.} In this work, we decided to take {a fiducial} scenario in which the mass constrictions given in Equation (\ref{m_22_300}) should work as lower and upper bounds for the mass parameter in the model, which turns out to be in agreement with {almost all of the constrictions presented above. The idea of taking these fiducial values is to be able to test throughout this work the different results we shall obtain. However, in section \ref{subsection42} we will actually test our results with observational data, so it will be in that section where we will actually constrain the mass parameter of the model.}}

\section{The SFDM model}\label{section3}

Now, {let us review the basics of} the SFDM model once more matter contributions are considered. Recall that the main goal of this work is to extend the results presented in section \ref{implicationB}. {For this purpose, in this section we will present the different equations that are necessary to describe galaxies in the SFDM model. After this, we will review an already known result for the solution of the soliton profile typically found in galactic centers: currently, the well-known mass-radius relation. We must mention that this result has only been reported when there are no other matter elements in the system, apart from the scalar field. For this reason, our intention in this section will also be to extend this result to the case in which the gravitational potential due to extra matter elements is incorporated, thus resulting in a more realistic representation of the system. Due to the difficulty of doing this numerically, we will adopt a Gaussian ansatz that will help us to be able to easily handle the free parameters of the system and thus achieve our mission. After this, we will review some results that have been found to be fulfilled by galactic halos in SFDM-only simulations (these relating the central soliton in halos to its host halo), to later try to use these relationships to extend the core-halo mass relation once we incorporate the contribution of extra elements of matter. For the latter, we will have to assume that the relationships that have been reported in SFDM-only simulations continue to be valid in this more general scenario.}

\subsection{Basic equations for the SFDM model}

As already mentioned, DM halos and their evolution along the structure-formation process in the SFDM model are described by classical solutions of the SP system \citep{wf1}:
\begin{subequations}\label{schpo}
\begin{equation}
    i\hbar \frac{\partial \psi}{\partial t}=-\frac{\hbar^2}{2m}\nabla^2 \psi+m\Phi \psi, 
\end{equation}
\begin{equation}\label{Poisson_new}
\nabla^2\Phi = 4\pi G\rho,    
\end{equation}
\end{subequations}
where $\rho$ is a cosmological overdensity that usually includes all of the contributions coming from DM, baryons, and, for example, the possible presence of a SMBH, that is, $\rho = \rho_{\psi}+\rho_{\rm{ext}}$, where $\rho_{\psi}=m|\psi|^2$ and $\rho_{\rm{ext}}=\rho_{\rm{bar}}+\rho_{\rm{SMBH}}$.  Equivalently, the previous system can be turned into its hydrodynamic representation by recasting the wave function $\psi$ in its polar form,
\begin{equation}
    \psi(\text{\textbf{r}},t) = \sqrt{\frac{\rho_{\psi}(\text{\textbf{r}},t)}{m}}e^{iS(\text{\textbf{r}},t)},
\end{equation}
and defining a velocity field as
\begin{equation}
    \bar v_\psi = \frac{\hbar}{m}\nabla S.
\end{equation}
Under this new set of variables, the SP system is rewritten as {an Euler-like system of differential equations}:
\begin{subequations}\label{hydroeq}
\begin{equation}\label{hydro_eq}
    \rho_\psi\frac{\partial \bar v_\psi}{\partial t}+\rho_\psi(\bar v_\psi\cdot \nabla)\bar v_\psi = -\rho_\psi \nabla Q-\rho_\psi\nabla \Phi,
\end{equation}
\begin{equation}\label{continuity}
    \frac{\partial \rho_\psi}{\partial t}+\nabla\cdot(\rho_\psi \bar v_\psi) = 0,
\end{equation}
\end{subequations}
where
\begin{equation}
    Q \equiv -\frac{\hbar^2}{2m^2}\frac{\nabla^2\sqrt{\rho_\psi}}{\sqrt{\rho_\psi}}.
\end{equation}
This $Q$ term is usually known as the quantum potential, and it arises from the quantum nature of SFDM. Additionally, in order to describe the properties and dynamics of SFDM halos, some physical quantities are needed, such as the total mass and total energy of the SFDM configuration, which can be computed from the solution as follows: 

\begin{equation}
    M = m\int_V |\psi|^2 d^3\text{\textbf{r}},
\end{equation}
and
\begin{equation}\label{equationE}
    E = \int_V\left[\frac{\hbar^2}{2m}|\nabla\psi|^2+\frac{m}{2}\Phi|\psi|^2\right]d^3\text{\textbf{r}}.
\end{equation}
Thus the total energy can be written as
\begin{equation}
    E = K + W,
\end{equation}
where 

\begin{equation}\label{equationK}
    K = \int_V\frac{\hbar^2}{2m}|\nabla\psi|^2d^3\text{\textbf{r}}  \end{equation}
is the total kinetic energy, and
\begin{equation}\label{equationW}
    W = \int_V\frac{m}{2}\Phi|\psi|^2d^3\text{\textbf{r}}
\end{equation}
is the total gravitational potential. These last expressions for the energy contributions are related by the scalar virial theorem of an isolated mass distribution, given by

\begin{equation}\label{virial}
    2K+W=0.
\end{equation}

\subsection{Soliton properties: quantum pressure vs gravity}

{It is now  accepted that at large time scales (structure-formation timescales) the averaged density profile of cores appearing in central regions of SFDM halos can be well fitted by coherent, quasi-stationary, ground-state solutions of the SP system}. These kinds of solutions are obtained once we assume a harmonic time dependence for the SFDM wavefunction $\psi$, or from the hydrodynamic representation, once we assume hydrostatic equilibrium ($\bar v_\psi = 0 = \partial \bar v_\psi/\partial t$). In the most-studied case, the source of the Poisson equation is given {only} by the scalar field density $\rho_\psi$, in which a {mass-radius relation} arises and is given by \cite{sf12}: 
\begin{equation}\label{mc_rc_num}
    R_{99} = 9.9\frac{\hbar^2}{GM_cm^2},
\end{equation}
where $R_{99}$ is the radius that contains $99\%$ of the total mass $M_c$ of the soliton profile.

{One of} our goals in this section is to extend the mass-radius relation for the soliton profile once non-SFDM  contributions are added to the system. Formally speaking, the SP system can only be solved numerically. However, for the sake of simplicity, it has been fitted with a functional form by \citet{sc4,sc10}, which turns out to be valid when  SFDM is the only contribution in the system (corresponding to the wave-SFDM density profile shown in Table \ref{tab:profiles}) {and which is based upon an empirical fit to the central region of simulated SFDM halos}; it also has been proposed to be approximated by a Gaussian distribution \citep{sc15,guzman2018head}. {The use of a Gaussian is motivated by the fact that Gaussian ``wave packets" not only appear in many contexts of a linear Schr\"odinger equation, but it also constitutes a solution for laboratory Bose-Einstein condensates; see, for example, \cite{BP}. In this work}, we adopt {to work with} the second option {over the first one given its better physical foundation and the fact that it is easier to find physical relations of interest from it. On the other hand, we decided to work with the Gaussian solution instead of the numerical solution given that once we consider more matter elements in the system, the numerical solution would be restricted to the distribution that we adopt for such extra matter elements. However, we will see later that the Gaussian ansatz will allow us to find general relationships without the need to adopt some profile for these extra elements of matter. Of course, the real solution must be the numerical one. However, it has been shown that this Gaussian ansatz can reproduce very well several of the results that have been found numerically for SFDM-only solitons, even if a self-interaction term is added to the system, a central black hole, or even if it is compared with results coming from general relativity, differing from these results only in the numerical factors that accompany each quantity, this discrepancy being in almost all cases $\sim O(1)$ \citep[for a better understanding of the latter, we encourage the reader to refer to][]{sc15,chavanis2019core,mipaper}. Thanks to these virtues of this Gaussian ansatz, we decided to proceed to work with it, accepting perhaps losing a bit of precision in the determination of the free parameters of the model (there should be discrepancies of the order $\sim O(1)$ in this determination), but acquiring the possibility of being able to describe in a simple way everything related to the soliton profile in the SFDM model.} 

The density profile for the soliton {with the Gaussian ansatz} is given by
\begin{equation}\label{deng1}
    \rho_c(r)= \frac{M_c}{(\pi R_c^2)^{3/2}}e^{-r^2/R_c^2},
\end{equation}
where the subindex $c$ stands for the ``core" quantities, and $R_c$ is a characteristic radius of the soliton profile, which is related to  $R_{99}$  as $R_{99}\simeq 2.38167R_c$. 
To achieve our purpose, we shall use the hydrodynamic representation of the SP equations. As we discussed previously, the soliton can be understood as a solution of {the SP system in its hydrodynamic representation} (\ref{hydroeq}) in the hydrostatic approximation:
\begin{eqnarray}
    \nabla Q &=& -\nabla \Phi,\label{hidrostatic}\\
    \frac{\partial\rho_\psi}{\partial t} &=& 0.
\end{eqnarray}

The second equation above follows from demanding hydrostatic equilibrium, while the first one reveals the physical {process} responsible for the formation of these solitons: in the free-field limit, they result from the  balance between the attractive gravity ($\nabla\Phi$ term) and the repulsion due to the uncertainty principle ($\nabla Q$ term). Notice that in Equation (\ref{hidrostatic}) the total gravitational potential $\Phi$ includes all of the contributions corresponding to all of the matter elements in the system. Consequently, the soliton profile that results {in this more general case arises from} the balance between the attraction due to the gravitational force produced by all of the matter elements and the repulsion provoked by the quantum nature of the SFDM particles.

After spherical symmetry {is imposed}, the gravitational force $\nabla\Phi$ can be calculated exactly as follows:

\begin{equation}
    \nabla\Phi(r) = \frac{GM_t(r)}{r^2},
\end{equation}
where $M_t(r)$ is the total mass (scalar field and extra constituents) enclosed inside a radius $r$. On the other hand, by using the Gaussian ansatz (\ref{deng1}), we get
\begin{equation}
    \nabla Q(r) = -\frac{\hbar^2}{m^2}\frac{r}{R_c^4}. 
\end{equation}
Using these last two expressions in (\ref{hidrostatic}) with $r = R_c$, we finally get the {extended mass-radius relation}
\begin{equation}\label{mass_radius_Rc}
 R_c = \frac{\hbar^2}{GM_t(r<R_c)m^2},
\end{equation}
which in terms of the radius that contains $99\%$ of the mass $M_c$ of the Gaussian ansatz reads as
\begin{equation}\label{mass_radius_r99}
   R_{99} = 2.38167R_c \simeq 32.17 \frac{\hbar^2}{GM_c^tm^2}.
\end{equation}
In the last equation, we have introduced $M_c^t\equiv M_t(r<R_{99})\simeq M_c+M_{\rm{ext}}(r<R_{99})$, and $M_{\rm{ext}}(r)$, which corresponds to the mass enclosed within radius $r$ corresponding to the extra matter constituents. Observe that the {extended mass-radius relation} differs from {the SFDM-only} exact numerical {mass-radius relation} (\ref{mc_rc_num}) for $M_c^t \simeq M_c$ by a factor of $\sim 3.2$. Nevertheless, this discrepancy is expected because we adopted an approximate solution to derive it. Although there is a discrepancy between the relations, given that the ratio of  both of them is $O(1)$, therefore within the context of our approximation the properties of solitons are quite similar than in the exact SFDM-only case. {Then, we assume that the above expression should also maintain the properties of solitons when more matter constituents are included in the total system.}

On the other hand, the Gaussian ansatz (\ref{deng1}) leads us to some other relations that will be relevant for this work. Notice that from the last equation and (\ref{deng1}) we obtain
\begin{equation}\label{mu_psi0}
\mu_\psi = \frac{1}{\pi^{3/2}R_c^2}\left[13.507\frac{\hbar^2}{Gm^2R_c}-M_{\rm{ext}}(r<R_{99})\right],    
\end{equation}
where we have defined a new surface-density parameter given by
\begin{equation}\label{mu_psii}
\mu_\psi \equiv \rho_sR_c= \frac{M_c}{\pi^{3/2}R_c^2}.
\end{equation}
This definition can be simply translated into the one we reviewed in section \ref{implicationB} {(specifically equation \eqref{surface_density_definition})}  by noticing that $r_s= R_{99}/3.77=2.38167 R_c/3.77$ (see next section).
We shall deal with this result in more  detail in section \ref{sec:IVA}. However, {at this point, it is interesting to observe that} $\mu_{\psi}$  maintains the same functional form  as Equation (\ref{mu_constant}) in the limit $M_{\rm{ext}}(r<R_{99})=0$.

\subsection{Extending the Core-Halo Mass Relation by Including the Baryonic Vontribution}\label{sectionIIIc}

Nowadays, our best understanding of a galaxy halo made up of an SFDM candidate can be well described by the results obtained in \cite{sc4,sc10}. {From these works, the averaged density profiles} of halos made of only SFDM particles {and at cosmological redshift $z=0$} can be well approximated by a {spherical} core-envelope structure, in which a central core transitions at a certain radius to an ``NFW-like" halo envelope. In that work, it was also realized that a core-halo mass relation exists, which in fiducial units reads as\footnote{{In fact, a different core-halo mass relation has also been reported in \cite{sc13}, in which the authors found that galaxies should harbor a more massive soliton. This last result was found by means of noncosmological, fully virialized settings. However, in this work, we decided to work with the results of \citet{sc4,sc10} given that these are more conservative for our purposes.}}:
\begin{equation}\label{M_c0}
    M_{s,7} \simeq 0.504\times 10^2 \frac{M_{h,12}^{1/3}}{ m_{22}},
\end{equation}
where $M_{s,7}\equiv M_s/(10^{7} M_\odot)$, and $M_s$ is the mass contained within radius
\begin{equation}\label{R_c}
    r_{s} = 1.6\times 10^{2}M_{h,12}^{-1/3}m_{22}^{-1}\ \rm{pc},
\end{equation}
defined as the radius at which the density of the central core drops by a factor of one-half from its value at $r=0$, and it is related to $R_{99}$ as $R_{99} \simeq  3.77\cdot r_{s}$, $M_{h,12}\equiv M_h/(10^{12}M_\odot)$, where subindex $h$ stands for ``halo"; in this case, $M_h$ is the total mass of an SFDM halo arising from  SFDM-only simulations. The core-halo mass relation can be expressed more conveniently in terms of quantities measured at $R_{99}$ instead of $r_s$ as done by \citet{wf2} as follows:
\begin{equation}\label{M_c}
    M_{c,7} \simeq 1.4\times 10^2 \frac{M_{h,12}^{1/3}}{ m_{22}},
\end{equation}
with $M_{c,7}\equiv M_c/(10^7 M_\odot)$. Then, the SFDM-only simulations for galaxy structures predict that, for a particular value of the halo mass $M_{h,12}$, there should exist a central soliton with a unique core mass $M_{s,7}$ (or equivalently $M_{c,7}$). This soliton may importantly determine the dynamics in bulges of large galaxies, either they would help to reproduce almost all of the total DM halos in small dSphs. Additionally, {the core-halo mass relation} (\ref{M_c}) predicts the minimum mass of a galactic halo within this model (i.e. when $M_h=M_c$). When the {fiducial} values (\ref{m_22_300}) for the mass parameter {are} used, the model predicts a minimum halo mass of $M_h^{min} \simeq (1.29\times 10^6 - 4.293\times 10^8)\ M_\odot$. Additionally, if the mass-radius relation (\ref{mc_rc_num}) is taken into account, the whole halo should have a total size of $R_h \simeq (4.71 - 32.55)\ \rm{kpc}$. {Interestingly, these low-mass and very large galaxies are a prediction for the SFDM model, which recently appears to be observed more and more in the universe. Some examples of these galaxies are the recent discovery Antlia I \citep{antlia1}, Crater II \citep{antlia2,antlia3}, and other large dSph galaxies in orbit around Andromeda \citep{antlia4}. Particularly, \cite{antlia5} it was tested the SFDM and obtained that a candidate with a mass of $m_{22}\sim 1.1$ is able to correctly fit the observations coming from Antlia I.}

{As we mentioned already, the core-halo mass relation was not anticipated by previous works. However, there have been several attempts to understand the physical meaning of {it} in an a posteriori manner}. On one hand, some authors \citep{chavanis2019core, chavanis2019predictive,sc13} have noticed that the above core-halo mass relation could be explained by assuming that the characteristic circular velocity at the core radius is roughly the same order as that at the halo radius {(``velocity dispersion tracing")}, implying that

\begin{equation}\label{vcvh}
    v_c\sim v_h \ \ \ \Rightarrow \ \ \ \ \frac{GM_c}{R_c}\sim \frac{GM_h}{R_h}.
\end{equation}
As explained in \cite{davies2020fuzzy}, {the meaning of the above relation is that the size of the soliton matches the de Broglie wavelength of the velocity dispersion $\sigma$ of the halo.} This equality is predicted from thermodynamics equilibrium arguments from kinetic theory {by equating $v_c\sim v_h\sim \sigma^2$ with the gravitational ``temperature" between the soliton and the halo}. On the other hand, {in \cite{wf2}, it was realized that the core-halo mass relation could be also understood as that the specific energy for the isolated central soliton and the total halo are the same, that is, that the condition 
\begin{equation}\label{ec_eh}
    \frac{|E_c|}{M_c}\simeq \frac{|E_h|}{M_h},
\end{equation}
applies{, where $E_c$ ($E_h$) is the total energy of the soliton (halo), and can be calculated with equation \eqref{equationE}}. Observe that, from the virial theorem (\ref{virial}), the above condition also implies the consequence
}
%
\begin{equation}\label{core_halo}
    \frac{K_c}{M_c}\simeq \frac{K_h}{M_h},
\end{equation}
{where, similarly, $K_c$ ($K_h$) is the total kinetic energy of the soliton (halo) and can be calculated with equation \eqref{equationK}.} An interesting feature of the last equations that is pointed out in a recent work \citep{core_halo1} is that the last relation (\ref{core_halo}) is more physically meaningful for explaining the core-halo properties of systems {than equation \eqref{ec_eh}, since they mention that relation \eqref{ec_eh} only applies when the quantities of an isolated soliton are compared and that such a relation should not really hold in the complete core-halo system, whereas Equation \eqref{core_halo} should still hold.\footnote{{We recommend to readers that if they want to delve further into this discussion, refer to \cite{core_halo1}. For the purposes we are interested in, this is not really important as we will show that all three relationships reduce to the same. This does not happen when other kinds of extra contributions are incorporated into the system, as it is the case of a self-interaction term between particles \citep[see, for example,][]{mipaper}.}}}

{Our second goal in this section is to extend the core-halo mass relation once more matter elements are incorporated into the system; that is, we are interested in finding the way in which the gravitational effects due to these extra elements of matter should affect equation \eqref{M_c0} (or equivalently equation \eqref{M_c}).} It is worth mentioning that so far the way in which baryons or other matter contributions affect the structure of virialized halos within the SFDM model is not fully understood, {although some analytical or simplified numerical settings have been proposed. For example, in \cite{wf2,core_halo1}, the SFDM model was studied in presence of baryons by assuming that Equation (\ref{ec_eh}) or (\ref{core_halo}) continues being valid. In a similar way, in \cite{chavanis2019core} it was proposed to extend the core-halo mass relation once including the presence of an SMBH was included and assuming that Equation (\ref{vcvh}) is still valid for the complete system}. Nevertheless, by virtue of the previous studies mentioned above, it is reasonable to assume that halos in the presence of more matter contents still hold a core-envelope structure, {fulfilling one of the previous relations}. {In fact, \cite{veltmaat2020baryon} \citep[see also][]{chan2018stars} studied the halo formation of SFDM including baryons and star formation, obtaining that this core-halo structure is maintained, although when they tried to prove if a relation of the form (\ref{core_halo}) still applied to the final halo formed, they found that there were some small deviations (up to an order of $\sim 2$) in which the equality in (\ref{core_halo}) was not fulfilled. However,} in despite of the simplicity of {our assumption}, it may bring up some insights that may be useful in describing in a more precise way the structure of halos in this more general regime and in figuring out possible scaling relations between different parts of this structure (as it happens in the SFDM-only case). {Then, in this section, we are interested in extending the core-halo mass relation for the SFDM once more matter constituents are added to the model and assuming that Equation (\ref{vcvh}), (\ref{ec_eh}), or (\ref{core_halo}) is valid.} {In fact, it is not difficult to realize that all three relations are the same, even when more matter components are added to the complete system.} We can reach such a conclusion by observing that by using the virial theorem (\ref{virial}) in (\ref{ec_eh}) and (\ref{core_halo}), the following relation can be obtained:
\begin{equation}\label{wc_wh}
    \frac{|W_c|}{M_c} = \frac{|W_h|}{M_h},
\end{equation}
{where, as before, $W_c$ ($W_h$) is the total gravitational potential energy for the soliton (halo) and can be calculated using equation \eqref{equationW}}. Second, the gravitational energy contributions  in this equation can be expressed as  
\begin{equation}
    |W_i|\sim \frac{GM_i M_i^t}{R_i}, \ \ \ \ \ i = c,h,
\end{equation}
By plugging the last relation into (\ref{wc_wh}), it turns into
\begin{equation}
    \frac{GM_c^t}{R_c}\sim \frac{GM_h^t}{R_h},
\end{equation}
where $M_h^t$ is the total mass  of the whole galactic system (SFDM, baryons, SMBH, and so on). Clearly, the last relation results from the assumption that the core and halo circular velocities are roughly equal even if extra matter sources are added to the galactic system. Or, in other words, the {previous relation should imply that the} coupled complete system of halo, soliton, and extra matter constituents should thermodynamically reestablish, from the kinetic theory arguments, the above new relation {such that, again, the size of the soliton still matches the de Broglie wavelength of the velocity dispersion $\sigma$ of the halo}. In this work, to achieve our purpose, we shall assume that the above expression is the one that is valid for explaining our extension\footnote{There is also a physical support to consider that the above expression is the one that is valid when there are more matter contributions: given the fact that DM interacts only gravitationally with the rest of the matter, it would allow us to suppose that the extension once more matter elements are considered would be to change quantities where the mass of the SFDM structures appears for the total mass (SFDM structure plus contributions of the other matter elements in the system).}. In fact, we shall use the relation 
\begin{equation}\label{vcvh_t}
    \frac{GM_c^t}{R_{99}}\simeq a\frac{GM_h^t}{R_h},
\end{equation}
with $a$ a constant of order $O(1)$ that will be fixed once compared with numerical simulations. {This parameter $a$ can be seen from the description that we have followed until now, for example, as a parameter that absorbs the errors that are added for our approximated Gaussian solution used to describe the soliton profile} \footnote{More generally but not considered in this work, this parameter could have the specific information on how DM interacts with baryons, and then it should not necessarily be a constant.}.  

Let us now extend our core-halo mass relation by including the effects of more matter contributions in the complete system. We can use the definition of the radius of the total galaxy of $R_h = R_{200}$, where $R_{200}$ is the radius at which the mean density inside such radius $\rho_{200}$ is $200$ times higher than the background density of the universe. This relation implies that $M_h^t = 4\pi \rho_{200}R_{h}^3/3$. If we replace this expression in (\ref{vcvh_t}) and use {the extended mass-radius relation} (\ref{mass_radius_r99}), we finally arrive at
\begin{equation}
    M_c^t \simeq \left(\frac{4\pi}{3}\right)^{1/6}\sqrt{\frac{32.2\, a\,\rho_{200}^{1/3}\,\hbar^2}{Gm^2}}\left(M_h^t\right)^{1/3},
\end{equation}
which can be also expressed in fiducial units as
{
\begin{equation}
    M_{c,7}^t \simeq \left(\frac{a^3\rho_{200}}{8.7\times 10^{-8}{M_\odot}/{\rm{pc}^3}}\right)^{1/6}\left(140 \frac{\left(M_{h,12}^t\right)^{1/3}}{m_{22}}\right).
\end{equation}
}
For the sake of consistency, the last expression must coincide with {the core-halo mass relation} (\ref{M_c}) in the SFDM-only limit, this is, when $M_i^t \simeq M_i$.  In that limit, the condition $a^3\rho_{200}= 8.651\times 10^{-8} M_\odot/\rm{pc}^3$ should be fulfilled\footnote{If we take the present time mean density of the universe $\rho_b =1.5\times 10^{-7} M_\odot/\rm{pc}^3$, we obtain that $a\simeq 0.66$.}. After taking into consideration the previous arguments, the final {extended core-halo mass relation that} turns out to be valid when more matter constituents are added to SFDM reads as
\begin{equation}\label{M_c_ext}
    M_{c,7}^t \simeq 1.4\times 10^2 \frac{\left(M_{h,12}^t\right)^{1/3}}{m_{22}}.
\end{equation}
This relation together with {the extended mass-radius relation} (\ref{mass_radius_r99}) leads to a characteristic core radius given by
\begin{equation}\label{r_c_ext}
    r_s \equiv \frac{R_{99}}{3.77} \simeq 5.21\times 10^2 (M_{h,12}^t)^{-1/3}m_{22}^{-1}\ \rm{pc}.
\end{equation}
Clearly, the discrepancy of the last relation between the core radius and Equation (\ref{R_c}) in the SFDM-only limit {is due to} the Gaussian approximation {used here}. Since we aim to elucidate some general predictions within the SFDM model arising from Equation (\ref{r_c_ext}), it is enough to assume that the last two relations roughly describe the core-halo mass scaling within the SFDM model.

It is interesting that {the extended core-halo mass relations} (\ref{M_c_ext}) and (\ref{r_c_ext}) have the same functional form as {the core-halo mass relations} (\ref{M_c}) and (\ref{R_c}), although they lead to different consequences. On one hand, both expressions imply that all galaxies have a  central core with a characteristic radii $r_s$ that systematically depend on the total mass of the galactic system. On the other hand, according to the extension derived here,  the amount of SFDM within this radius would be affected by the presence of extra matter constituents {because the total mass is the one that scales with $r_s$}. Therefore,
\begin{equation}
    M_{c,7} \simeq 1.4\times 10^2 \frac{\left(M_{h,12}^t\right)^{1/3}}{m_{22}}-M_{\rm{ext},7}, 
\end{equation}
where we have defined  $M_{\rm{ext},7}\equiv M_{\rm{ext}}(r<R_{99})/(10^7 M_\odot)$. Moreover, our extension predicts that, once the total halo mass $M_{h,12}^t$ is given, the amount of SFDM in the galactic central region has to be smaller than that predicted in the SFDM-only simulations owing to the contribution of extra matter components. {It should be stressed that a further analysis is needed in order to verify in a more precise way these scaling relations within the cosmological context because full-numerical cosmological simulations of structure formation are still lacking. Nevertheless, in despite of their simplicity, they bring some insight into more realistic systems}. {For instance,} {it is intriguing} that our humble result is able to predict a small amount of SFDM in galactic nuclei of large galaxies, which turns out to be in accordance with the fact that large galaxies typically contain a large amount of baryons in their centers. This point will be further explored in the next section, where we shall show that this realization is in accordance with the observations and results of McGaugh when applied to the SFDM context.

\section{Consequences in the SFDM from the McGaugh correlation function}\label{section4}

\subsection{General considerations}\label{sec:IVA}
In this section, {our goal is to compare} the SFDM model against the McGaugh observational correlations described in section \ref{sectionII},  {but now with our new extended relations}. Afterward, we shall extend the results we obtained in Section \ref{implicationB} {beyond the scope of dSphs}. 
With that purpose, we shall focus on the central region of SFDM halos where presumably the soliton/core is located \citep[see, for example,][]{veltmaat2020baryon,chan2018stars}. {Such a selection is done in order to remain in accordance with the semianalytical results discussed in Section \ref{sectionIIIc}, which are to be verified and precised once cosmological simulations with baryons are carried out.}

From the definition of the observed acceleration $\nabla\Phi(r) = g_{\rm{obs}}$, together with Equation (\ref{hidrostatic}), the following interesting relation for $g_{\rm{obs}}$ arises:
\begin{equation}
    g_{\rm{obs}} = -\nabla Q.
\end{equation}
This suggests that the quantum potential is intimately related to the acceleration observed in stars in the center of SFDM halos. Additionally,  the acceleration produced by the SFDM in the soliton can be recast by using the Gaussian ansatz, leading to
\begin{equation}
    g_{\psi} = {4\pi G \mu_{\psi}}\left[\frac{1}{4x^2}\left(\sqrt{\pi}Erf(x)-2xe^{-x^2}\right)\right],
\end{equation}
where $Erf(x)$ denotes the error function and $x \equiv r/R_c$.  The acceleration produced by the soliton profile turns out to be maximum at 
\begin{equation}
    g_{\psi,max} \simeq 3.34\times 10^{-11} \frac{\mu_\psi}{M_\odot \rm{pc}^{-2}} \frac{\rm{m}}{\rm{s}^2}.
\end{equation} 
Thus, notice that the maximum possible acceleration of SFDM particles at the galactic center only depends on the value of the surface density parameter $\mu_\psi$, as anticipated in section \ref{sectionII} {(equation \eqref{mu_const} and the discussion that follows from it)}. Furthermore, in that same section {(equation \eqref{g_maximum})}, it was realized that the maximum  acceleration produced by any candidate particle of DM satisfies $g_{h,max} = 0.65 g^{\dagger}$. If we match this value with that corresponding to the soliton (this last condition holds at least for dSphs, where the whole halo corresponds only to a soliton), we obtain (from {equation} \eqref{mu_const})
{ \begin{equation}\label{mu_psi}
    \mu_{\psi} \simeq 2.33\times  10^2 \frac{M_\odot}{\rm{pc^2}}.
\end{equation}}
It is important to stress that this result is valid for galaxies where a maximum possible acceleration produced by the SFDM particles occurs inside the soliton region; {it does not matter whether} this acceleration {takes the maximum value} in other parts of the halo {as long it occurs at least once inside the core}. In addition, as mentioned {in the introduction of this paper}, a host of  observations  \citep{scalingmu1,spano2008ghasp,donato2009constant,scalingmu2} suggest that the surface density appears to be constant for any type of galaxy. The previous argument together with this observational realization imply that the maximum acceleration in SFDM solitons and that inferred from the observations {of McGaugh et al.} must be always equal. In other words, independent of the amount of extra matter constituents contained in the central region of galaxies, we must have from the McGaugh results (\ref{g_h_max}) that $\mu_{\psi}$ must be a constant quantity in all galaxies and that the maximum possible acceleration produced by the SFDM happens inside the soliton region. If this were not the case, we would have that $\mu_{\psi}$ would decrease. Thus, eq.(\ref{mu_psi}) represents also the maximum possible value of the surface density within this model.

Let us {study more closely} Equation (\ref{mu_psi}). First of all, observe that (\ref{mu_psi}) implies from Equation (\ref{mu_psii}) a condition for the soliton mass $M_{c,7}$ in terms of the characteristic radius size $R_c$:
\begin{equation}\label{mr-soliton}
    M_{c,7} = 1.298\times 10^{-4}\left(\frac{R_c}{\rm{pc}}\right)^2.
\end{equation}
Besides, by using (\ref{mu_psi}) and (\ref{mu_psi0}), in fiducial units, we get
\begin{equation}\label{constr_r_c}
    \frac{1.12}{10^9}\simeq \left(\frac{R_c}{\rm{pc}}\right)^{-3}\left[m_{22}^{-2}-\frac{8.66}{10^{6}} M_{\rm{ext},7}\left(\frac{R_c}{\rm{pc}}\right)\right].
\end{equation}
In a similar way, by plugging {the extended mass-radius relation} (\ref{mass_radius_r99}) into (\ref{mu_psii}) and then into (\ref{mu_psi}), we obtain
\begin{equation}\label{mcmt2}
 M_{c,7}(M_{c,7}^t)^2 \simeq \frac{1.636\times 10^5}{m_{22}^4}.
\end{equation}
{It is not difficult to realize that these last three relationships must also be fulfilled by those galaxies that do not necessarily fulfill the McGaugh et al. results but that do maintain the same constant value of the surface density for this model. }

{For the results presented in this section to be valid, a necessary condition that must be met is that in the limit where all extra matter elements can be ignored in the system, we should reproduce all of the results presented in Section \ref{implicationB} for the SFDM model.} Notice that when $M_{\rm{ext},7}\ll M_{c,7}$, which would apply for typical dSphs, the above two expressions lead to
\begin{subequations}
\begin{equation}\label{m_R_c}
   \left(\frac{R_c}{\rm{pc}}\right)^{-3}m_{22}^{-2} \simeq 1.12\times 10^{-9}, 
\end{equation}
and
\begin{equation}\label{m_M_c}
    M_{c,7}\simeq \left(\frac{1.636\times 10^5}{m_{22}^4}\right)^{1/3},
\end{equation}
\end{subequations}
The last two expressions imply that halos of this kind of galaxies have a central soliton with nearly the same radius and mass (with very small deviations due to the baryonic contribution). {Consequently,}  the value of the total mass enclosed within a given radius must be nearly a constant. This conclusion is {similar to} that obtained in section \ref{implicationB} derived from SFDM-only simulations, {which gives us confidence in the results obtained from these more general relations \eqref{constr_r_c} and \eqref{mcmt2}}. Even more, the dependence of the SFDM particle mass and the core radius (see equations \eqref{mu_constant} and \eqref{m_R_c}) remains to be quite similar. As expected, the numerical values involved are different because the last two equations were obtained from the Gaussian ansatz. {For completeness and consistency, we estimate the value of the SFDM mass parameter following the same prescription used to derive  Equation (\ref{m_22_300}) but using results derived from the Gaussian ansatz: }
\begin{equation}\label{m_gaussian}
    m_{22}\simeq 0.55 \ \ \ \text{and} \ \ \ m_{22}\simeq 23.31.
\end{equation}
Of course, as we already discussed, these mass parameters must be refined once we directly compare this model with the data coming from dSphs of the Milky Way.
Observe that although these parameters involve slightly different numerical values than the ones obtained in Equation (\ref{m_22_300}), neither differs too much, which proves that our approximation turns out to be very good. For consistency in description in this section, we will now move on to treating the previous mass parameters as our new fiducial parameters for the model, similar to what we previously did with the parameters in equation \eqref{m_22_300}.

\begin{figure}[t]
    \includegraphics[width=0.45\textwidth]{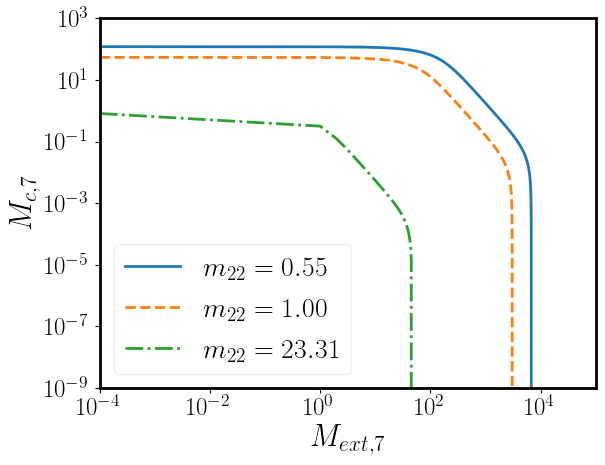}
    \includegraphics[width=0.45\textwidth]{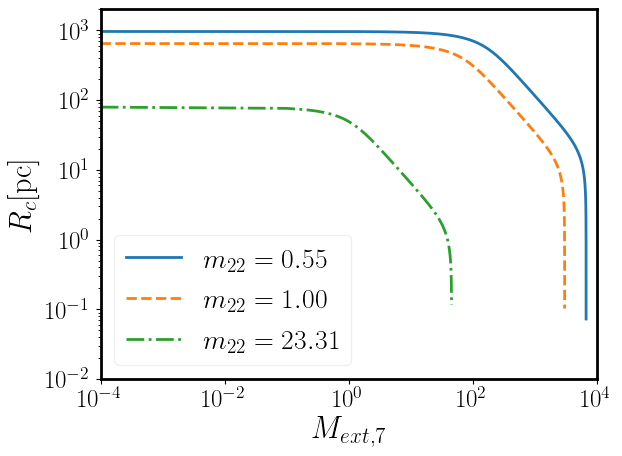}\\
    \includegraphics[width=0.45\textwidth]{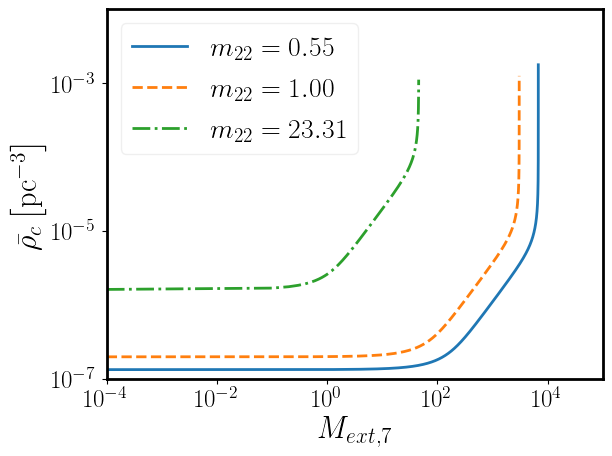}
    \caption{Top:$M_{c,7}$, middle:$R_{c}$, and bottom: $\bar\rho_c \equiv M_{c,7}/R_c^3$  as a function of $M_{\rm{ext},7}$ for three different values of the SFDM particle mass.}
    \label{fig:mcmt2}
\end{figure}

It is not difficult to realize that the conclusion of {the existence} of a core with universal mass and core-radius values may not hold for all types of galaxies because, in the most general {constraints} (\ref{constr_r_c}) and (\ref{mcmt2}), both quantities depend on the amount of other components different from SDFM contained within the soliton region. We show this dependency in Figure \ref{fig:mcmt2}, where we plotted the variation of $M_{c,7}$ (top), $R_c$ (middle), and $\bar\rho_c \equiv M_{c,7}/(R_c^3)$ (bottom) in terms of the mass $M_{\rm{ext},7}$ contained within the soliton region for three different values of $m_{22}=0.55,\ 1,\ 23.31$. As expected and as we can observe from the figure, when $M_{\rm{ext},7}$ is small, $R_{c}$ and $M_{c,7}$ remain approximately constant and equal to Equations(\ref{m_R_c}) and (\ref{m_M_c}); as a result, $\bar\rho_c$ remains constant as well. As $M_{\rm{ext},7}$ begins to grow, the characteristic radius $R_c$ must begin to decrease in order to fulfill with Equation (\ref{constr_r_c}), which would agree with assuming that the contribution of extra matter elements in the central region of galaxies would tend to compress the central soliton in galaxies. Since Equation (\ref{mr-soliton}) must be fulfilled as well, this implies that $M_{c,7}$ ($\bar\rho_c$) should start decreasing (increasing) as well. Particularly, the mass at which the mass produced by extra contributions equals the soliton mass $M_{\rm{ext},7}\simeq M_{c,7}^t/2\simeq M_{c,7}$ is
\begin{equation}
    M_{c,7}^t\simeq \left(\frac{3.272\times 10^5}{m_{22}^4}\right)^{1/3},
\end{equation}
or equivalently
\begin{subequations}
\begin{equation}\label{M_l}
    M_{c,7}\simeq M_{\rm{ext},7}\simeq \left(\frac{4.09\times 10^4}{m_{22}^4}\right)^{1/3},
\end{equation}
with a characteristic core radius:
\begin{equation}\label{R_l}
    R_{c}\simeq \frac{5.152\times10^2}{m_{22}^{2/3}}\ \rm{pc}.
\end{equation}
\end{subequations}
If we use the {fiducial} limited masses in (\ref{m_gaussian}), these last two quantities result:
\begin{equation}
    M_{c,7}\simeq 76.458, \ \ \ \text{and} \ \ \ M_{c,7} \simeq 0.517, 
\end{equation} 
with the characteristic core radius
\begin{equation}
    R_c\simeq 767.49\ \rm{pc},\ \ \ \text{and}\ \ \ R_c\simeq 63.13\ \rm{pc}.
\end{equation}
Then, for such galaxies with a mass $M_{\rm{ext},7}$ equal to or larger than Equation (\ref{M_l}) within radius  (\ref{R_l}) or smaller, it must be expected that the central region in galaxies should be dominated by these extra contributions. Finally, for a particular mass $M_{\rm{ext},7}$, which is given by
\begin{equation}
    m_{22}^{-2}\simeq \frac{8.66}{10^{6}} M_{\rm{ext},7}\left(\frac{R_c}{\rm{pc}}\right), \nonumber
\end{equation}
$R_c$ and $M_{c,7}$ decay abruptly to zero. This relation implies that for such galaxies that fulfill the above condition, they must not have a soliton/core in their center. 

\begin{figure}
    \includegraphics[width=3.2in]{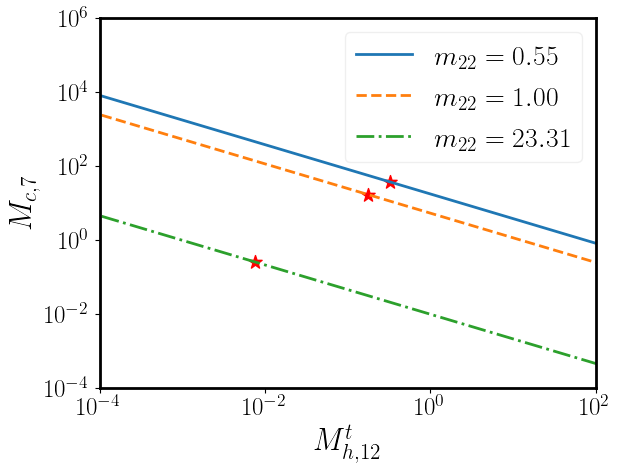}
    \caption{$M_{c,7}$ as a function of $M_{h,12}^t$ coming from Equation (\ref{eq:masa_core_halo}) for three different values of the SFDM particle mass. The red stars show the points where the mass of the central core is equal to the mass of the extra matter constituents.}
    \label{fig:mcmh}
\end{figure}

We have to stress out that, until now, we have not used our { extended core-halo mass relation} (\ref{M_c_ext}), and then these results are general as long as it is assumed that in the central region of galaxies there should exist a solitonic core described by the minimum-energy, coherent, quasi-stationary solution of the SP system. However, very intriguing is the fact that in our extension (\ref{M_c_ext}) as well as in the consequences of the McGaugh results (\ref{mcmt2}) we would have that the mass of the core would be diminished as long as the mass contribution of other matter components increases, a result that is expected to occur in the most massive galaxies but that is not obtained in SFDM-only simulations\footnote{In fact, in SFDM-only simulations, exactly the opposite consequence is expected. Then the results coming from SFDM-only simulations are in disagreement with the McGaugh et al. results.},  (see {the core-halo mass relation} \ref{M_c}). In fact, if we replace our {extended core-halo mass relation} (\ref{M_c_ext}) in (\ref{mcmt2}), we must have that

\begin{equation}\label{eq:masa_core_halo}
    M_{c,7} \simeq  \frac{5.266}{m_{22}^2}(M_{h,12}^t)^{-2/3}.
\end{equation}
We show in figure \ref{fig:mcmh} $M_{c,7}$ as a function of $M_{h,12}^t$ for the same mass parameters we used in figure \ref{fig:mcmt2}. Observe that we have also plotted the mass at which the mass of the central core equals the mass of the extra matter constituents, that is, when
\begin{equation}
    M_{h,12}^t \simeq \frac{1.782\times10^{-1}}{m_{22}}.
\end{equation}
Of course, as we already mentioned, the results obtained from our extension should be refined by means of cosmological simulations of galaxy formation in the presence of more matter constituents in the complete system. However, observe that this description also implies a characteristic halo mass at which the central region of galaxies must start to be dominated by extra matter contributions. Additionally, we can also see that this description can also alleviate the claim that the SFDM model is not capable of reproducing profiles with constant central surface density. 
\subsection{Constraining the SFDM Mass Parameter from Observational Evidence}\label{subsection42}

In this section, we are interested in confronting all of the conclusions that we have obtained until now. It is clear that our results should correctly describe typical dSphs. However, our intention here is to show the predictive ability that the SFDM model would have, and then, for our purpose, we decided to use data from another kind of source. Particularly, in order to constrain its mass parameter $m_{22}$, we decided to use observational data from the MW galaxy, {which is a spiral galaxy like those that are well known to meet McGaugh's relationship,} and following the same treatment of the contributions of matter in the MW as in \cite{maleki2020constraint}. The procedure we take in this section is then the following. First, we fit the rotation curves for stars in the MW once we assume the galactic halo is a central soliton, described by the Gaussian ansatz, and an NFW-like envelope structure. For this, the only constraint we took in the halo made of SFDM was to impose continuity at the match point between the soliton and the NFW exterior. Posteriorly, by assuming that the McGaugh observational correlation that led us to Equation (\ref{mcmt2}) is valid,  we found the value of $m_{22}$ that should match with that result. That is, we can assume that $\mu_\psi$ has the value shown in Equation (\ref{mu_psi}), coming from the McGaugh results. In such case, Equation (\ref{mcmt2}) should hold, and from there we can find the value of $m_{22}$ (shown in Equation \eqref{mass_IVB1}). Finally, we tested (\ref{eq:masa_core_halo}) with our results, which would end up being our most restrictive relationship because it assumes that Equation (\ref{mcmt2}) is valid, as well as our core-halo extension. With this equation, we find the value of the mass of the SFDM (\ref{mass_IVB2}) that would fulfill this relationship. Given our procedure, let us present each of the steps mentioned above in more detail.

We model the MW by considering a  stellar disc, a central bulge, a DM halo, and a central SMBH. As explained in \cite{maleki2020constraint}, in reality, the system is more complicated because it contains more contributions, as it is the case of the arms and bars of the galaxy, but these features can be considered as second-order terms without considerable effects \citep{effects1,MWdata}. With all these components, the total circular velocity of a particle at radius $r$ can be expressed as
\begin{equation}
    v(r) = \sqrt{v_h^2(r) + v_{d}^2(r) + v_{\rm{BH}}^2(r) + v_{b}^2(r)},
\end{equation}
where the subscripts ($h,BH,d,b$) stand for halo, black hole, disc and bulge, respectively. 

\textit{The black hole.} This can be modeled as a point-like object in the galaxy center. The circular velocity of a particle at radius $r$ due to the SMBH is 
\begin{equation}
    v_{\rm{BH}} = \sqrt{\frac{G M_{\rm{BH}}}{r}},
\end{equation}
where $M_{\rm{BH}}$ is the mass of the SMBH. 

\textit{The galactic disc.} The disc is modeled with a razor-thin exponential surface mass density of the form
\begin{equation}
    \Sigma_d (\rho) = \Sigma_0 e^{-\rho/a_d},
\end{equation}
where $a_d$ is the disc scale length and $\rho$ the cylindrical coordinate. The surface density at $\rho=0$, $\Sigma_0$, is linked to the total mass of the disc $M_d$ as $M_d = 2\pi\Sigma_0 a_d^2$. As derived from \cite{freeman1970ApJ}, the circular velocity due to this density profile is 
\begin{equation}
    v_{d} = \sqrt{\frac{G M_d y^2}{2 a} \left( I_0\left(\frac{y}{2} \right) K_0\left(\frac{y}{2} \right) - I_1\left(\frac{y}{2} \right) K_1\left(\frac{y}{2} \right) \right)}.
\end{equation}
Here, $I_n$ and $K_n$ are the modified Bessel functions of the first and second kind, respectively, and $y\equiv r/a_d$. 

\textit{The bulge.} This is modeled by de Vancoulers's suggestion \citep{vancou1,vancou2}, in which the bulge can be expressed as an exponential density of the form
\begin{equation}
    \rho_b (r) = \rho_0 e^{-r/a_b},
\end{equation}
where $a_b$ is the bulge scale length and the central density $\rho_0$ is linked to the total mass of the bulge $M_b$ as $\rho_0 = M_b/(8 \pi a_d^3)$. The respective circular velocity due to this profile is
\begin{equation}
    v_b = \sqrt{\frac{G M_b}{r} \left( 1 - \left(1 + \frac{r}{a_b} + \frac{r^2}{2a_b^2} \right)e^{-r/a_b}\right) }.
\end{equation}
It is well known that this profile is not able to explain the MW's rotation curve in the central region, and then the existence of a second bulge is usually assumed, necessary to correctly match the observational data. However, as shown in \cite{maleki2020constraint}, it is not necessary to introduce this second bulge because it can be explained by the presence of the central soliton present in the SFDM model. Therefore, in this section we adopt working in this way. 

\textit{The halo.} We assume the DM halo as a Gaussian core with a density defined as in Equation (\ref{deng1}) with an NFW-envelope exterior:
\begin{equation}
    \rho_{\rm{NFW}}(r)= \frac{\rho_{sN}}{\frac{r}{r_{sN}}\left(1 + \frac{r}{r_{sN}}\right)^2},
\end{equation}
where $\rho_{sN}$ and $r_{sN}$ are the NFW characteristic density and radius, respectively.
Then, the total density of such a halo can be written as
\begin{equation}
    \rho(r) = \theta (r_e-r) \rho_{c}(r) + \theta (r - r_e)\rho_{\rm{NFW}}(r)
\end{equation}
where $\theta$ is the Heaviside function and $r_e$ is the radius where the transition from soliton core to NFW envelope occurs. If we demand the density to be continuous at the transition radius, $\rho_{c}(r_e)= \rho_{\rm{NFW}}(r_e)$, we can reduce by one the number of parameters by expressing $\rho_{sN}$ in terms of the rest of the parameters as
\begin{equation}
    \rho_{sN} = \frac{M_c}{\pi^{3/2}R_c^3} e^{-\frac{r_e^2}{R_c^2}}\frac{r_e}{r_{sN}}\left(1 + \frac{r_e}{r_{sN}}\right)^2.
\end{equation}
The DM mass enclosed at radius $r$ is
\begin{eqnarray}
     M_h(r) &=& \theta (r_e-r) M_{core}(r)+\theta (r-r_e) f(r_e) \nonumber \\
     &+&  \theta (r-r_e) (M_{\rm{NFW}}(r)-M_{\rm{NFW}}(r_e)) \nonumber
\end{eqnarray}
where we have defined
\begin{eqnarray}
 M_{core}(r) &=& \frac{M_c}{\sqrt{\pi } R_c}\left(\sqrt{\pi } R_c \text{erf}\left(\frac{r}{R_c}\right)-2 r e^{-\frac{r^2}{R_c^2}}\right), \nonumber \\
 M_{\rm{NFW}}(r) &=& 4 \pi  \rho_{sN} r_{sN}^3 \left( \frac{1}{1 + \frac{r}{r_{sN}}}+\log \left(1+\frac{r}{r_{sN}}\right)\right).\nonumber
\end{eqnarray}
From $M_h$ we can find the circular velocity of a particle at a radius $r$ due to the DM halo:
\begin{equation}
    v_h(r)= \sqrt{\frac{G M_h}{r}}.
\end{equation}

Summarizing, for this model we have a total of nine parameters to fit, namely $(R_c, M_c, r_e, r_{sN}, M_d, a_d, M_b, a_b, M_{\rm{BH}})$. We fit the rotation curve data of \cite{MWdata} using the Markov Chain Monte Carlo (MCMC) method, sampling the parameter space from uniform priors in the range shown in Table \ref{tab:fitMW} \citep[for a review of parameter inference with MCMC techniques, see][]{padilla2019cosmological}.

\begin{figure}
    \centering
    \includegraphics[width=3.6in]{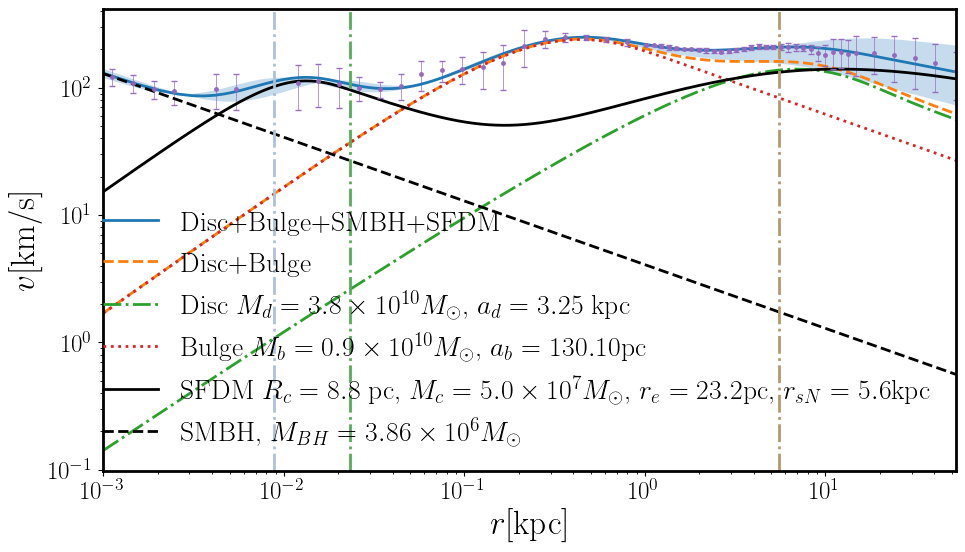}
    \caption{Milky Way rotation curve. The blue continuous line is the MCMC fit to the data, and the shaded region represents a 2$\sigma$ spread of the fit parameters. The contributions of the disc, bulge, SFDM halo, and central SMBH are also shown separately. The vertical lines from left to right correspond to the values of $R_c, r_e$ and $r_{sN}$ respectively. Data with error bars are from \citep{MWdata}.}
    \label{fig:fitMW}
\end{figure}

\begin{figure*}
    \centering
    \includegraphics[width=7in]{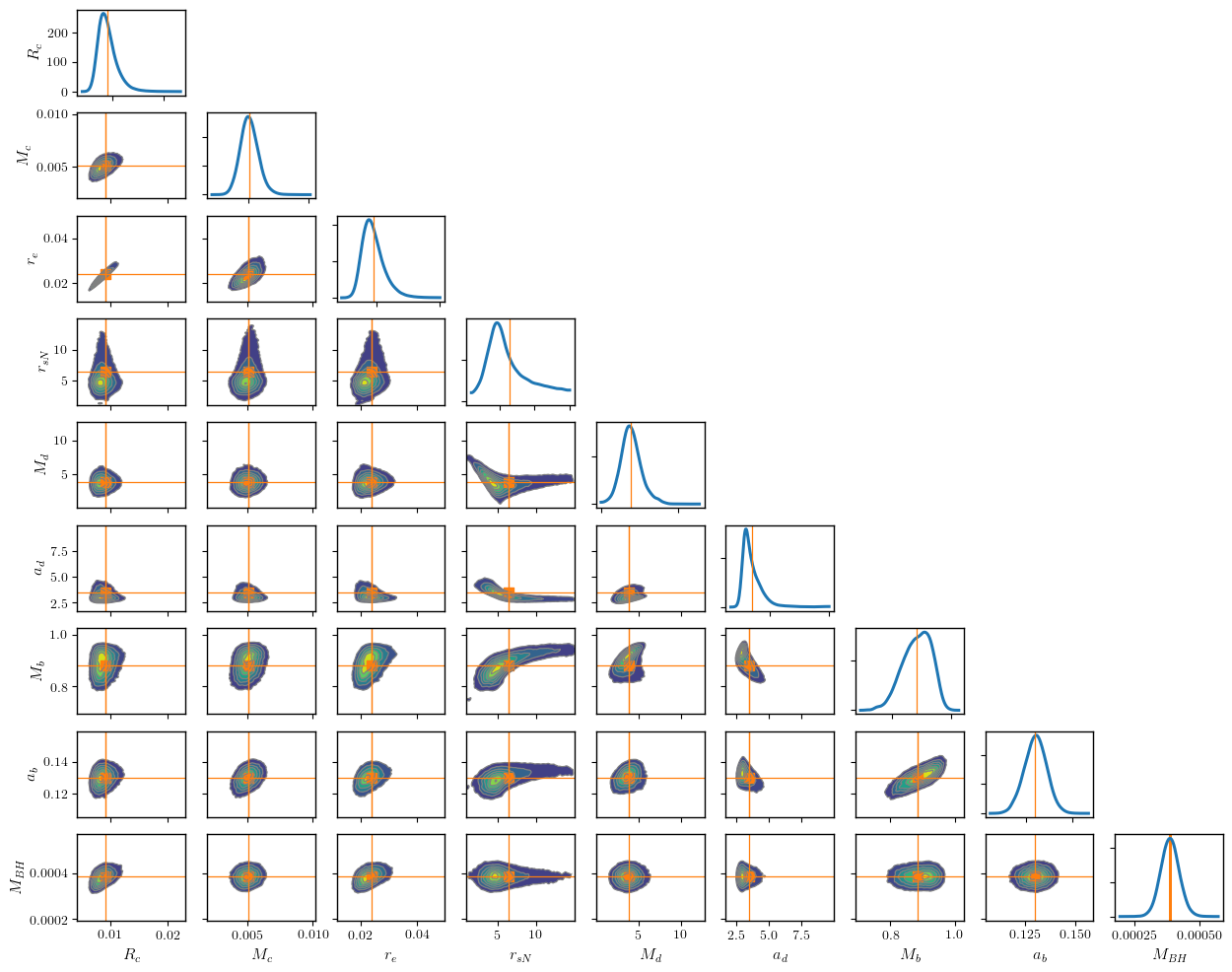}
    \caption{Milky Way MCMC fit posterior parameters. Here, $R_c, r_e, r_{sN}, a_d, a_b$ are in kpc units and $M_c, M_d, M_b, M_{\rm{BH}}$ in $10^{10}M_\odot$ units. Orange lines are located at the mean of the distribution.}
    \label{fig:fitMWparams}
\end{figure*}

The posterior parameters and the $1\sigma$ and $2\sigma$ confidence levels are calculated using the Lmfit and Emcee Python packages, and they are shown also in Table \ref{tab:fitMW}. Additionally, the fit of the rotation curves of the MW with our estimated parameters is shown in figure \ref{fig:fitMW} and the posterior distribution trace plots are shown in figure \ref{fig:fitMWparams}.

\begin{deluxetable}{lcccc}
\tablecaption{Priors, Mean Value, and 1$\sigma$ and 2$\sigma$ Spread of the Parameters of the MCMC Fitting of the Milky Way\label{tab:fitMW}}
\tablewidth{0pt}
\tablehead{\colhead{Parameter} & \colhead{Prior} &
\colhead{Mean}& \colhead{1$\sigma$} & \colhead{2$\sigma$}}
\decimalcolnumbers
\startdata
$R_c$ (pc) &[1,100]& 8.83   & 1.61   & 3.54   \\ 
$M_c$ ($10^7 M_\odot$)  &[1,50] & 4.99   & 0.73   & 1.51   \\ 
$r_e$ (pc) & [10,50] & 23.24  & 4.00   & 8.68   \\ 
$r_{sN}$ (kpc) & [1,15] & 5.55   & 3.28   & 6.10   \\
$M_d$ ($10^{10} M_\odot$) & [0.001,30] & 3.81   & 1.24   & 2.80   \\ 
$a_d$ (kpc) &[0.1, 10]& 3.25   & 0.68   & 1.78   \\ 
$M_b$ ($10^7 M_\odot$)  & [100,1500]& 887.13 & 49.48  & 89.92  \\
$a_b$ (kpc) & [0.05,0.5]& 0.13   & 0.006   & 0.01   \\ 
$M_{\rm{BH}}$($10^6 M_\odot$)  & [1,10]& 3.86   & 0.35   & 0.70   \\ 
\enddata
\end{deluxetable}

Once we have the best parameters for our model, we can test if Equation (\ref{mcmt2}) holds. We calculate the values of $M_{c,7}$ and $M_{c,7}^t$ and we obtain from Equation (\ref{mcmt2}) the value that the mass of the SFDM candidate should have to fulfill with such condition: 
\begin{equation}\label{mass_IVB1}
    m_{22} = 5.49 \pm 5.10.
\end{equation}
If finally we assume relation (\ref{eq:masa_core_halo}) holds, then we obtain
\begin{equation}\label{mass_IVB2}
    m_{22} = 1.41 \pm 1.07. 
\end{equation}
Observe that these two mass parameters agree with each other. Additionally, they agree with our {fiducial region of parameters. Even more important, if we compare with the constraints previously reported for the model, coming from other types of observations (see the penultimate paragraph of section \ref{section2}), we can see that our results also agree with most of them.}

\section{Conclusions}\label{section5}

In this work, we consider a Gaussian profile as an approximate solution of the Schr\"odinger-Poisson system, and then we assume that at distances of the size of the core, the components of matter can be considered spherically symmetrical. Using this approximation, we present an extension of the mass-radius relation by taking into consideration extra matter constituents of the galaxy, such as baryonic matter and a central SMBH. With this relation and the result of a constant central surface DM density arising from the McGaugh radial acceleration relation, we derived a constraint for the SFDM mass parameter, given by $m = (5.49 \pm 5.10) \times 10^{-22}\rm{eV}/c^2$. The method we use to compute this estimate was the following. First, by using an MCMC method, we determine the parameters that provide the best fit to the data of the circular velocity of stars at radii from a few parsecs to almost hundreds of kiloparsecs from the center of the Milky Way. Second, we use those parameters to calculate the total mass enclosed at a radius $R_{99}$, and finally, from there and Equation (\ref{mcmt2}), we obtained $m$. 

In the same way, by using the Gaussian approximation for the density profile of a soliton and assuming that the circular velocities at the core and at the halo are roughly the same, we can also extend the core-halo mass relation for the case when extra matter constituents are taken into consideration. Thus, applying our extension to our result of a constant central surface DM, we obtained our more restrictive conclusion given by Equation (\ref{eq:masa_core_halo}). When we compare Equation (\ref{eq:masa_core_halo}) with the constrictions coming from the MW, we obtain an estimate for the particle mass of $m = (1.41 \pm 1.07) \times 10^{-22}\rm{eV}/c^2$.

Taking into account both previous results, we obtain a mass with an order of magnitude of $10^{-22}\rm{eV}/c^2$ as the most preferred by observational data. Our final constrictions, Equations (\ref{mass_IVB1}) and (\ref{mass_IVB2}), turn out to be consistent at $1 \sigma$ with each other and with most of the cosmological constraints. The deviation of these two values could come from the approximations made to derive each of the relations, namely, the Gaussian ansatz and the assumed spherical symmetry in the central region of the galaxy or from nonthermal equilibrium between the core and the halo. These results encourage further numerical studies that take into account the baryonic contributions in the central regions of the galaxies. 

\acknowledgments
L.P. acknowledges sponsorship from CONACyT through grant CB-2016-282569. J.S. acknowledges financial support from a CONACyT doctoral fellowship. This work was partially supported by CONACyT M\'exico under grants 240512, A1-S-8742, 304001, 376127;
Xiuhcoatl and Abacus clusters at Cinvestav, IPN; and
I0101/131/07 C-234/07 of the Instituto
Avanzado de Cosmolog\'ia (IAC) collaboration (http://www.iac.edu.mx/). This research received support from Conacyt through the Fondo Sectorial de Investigaci\'on para la Educaci\'on, grant No. 240512.

\vspace{5mm}
\software{Lmfit Python package \citep{lmfit},
          Emcee Python package \citep{emcee}}

\bibliography{references}{}

\begin{thebibliography}{}
\expandafter\ifx\csname natexlab\endcsname\relax\def\natexlab#1{#1}\fi

\bibitem[{Arbey {et~al.}(2001)Arbey, Lesgourgues, \& Salati}]{sf6}
Arbey, A., Lesgourgues, J., \& Salati, P. 2001, Physical Review D, 64, 123528

\bibitem[{Arbey {et~al.}(2002)Arbey, Lesgourgues, \& Salati}]{sf5}
---. 2002, Physical Review D, 65, 083514

\bibitem[{Arbey {et~al.}(2003)Arbey, Lesgourgues, \& Salati}]{sf7}
---. 2003, Physical Review D, 68, 023511

\bibitem[{Armengaud {et~al.}(2017)Armengaud, Palanque-Delabrouille, Y{\`e}che,
  Marsh, \& Baur}]{free_const6}
Armengaud, E., Palanque-Delabrouille, N., Y{\`e}che, C., Marsh, D.~J., \& Baur,
  J. 2017, Monthly Notices of the Royal Astronomical Society, 471, 4606

\bibitem[{Avilez \& Guzm\'an(2019)}]{Avilez-Lopez:2018hwh}
Avilez, A.~A., \& Guzm\'an, F. 2019, Phys. Rev. D, 99, 043542

\bibitem[{Avilez {et~al.}(2018)Avilez, Padilla, Bernal, \&
  Matos}]{Avilez-Lopez:2017zfp}
Avilez, A.~A., Padilla, L.~E., Bernal, T., \& Matos, T. 2018, Mon. Not. Roy.
  Astron. Soc., 477, 3257

\bibitem[{Bar {et~al.}(2018)Bar, Blas, Blum, \& Sibiryakov}]{wf2}
Bar, N., Blas, D., Blum, K., \& Sibiryakov, S. 2018, Physical Review D, 98,
  083027

\bibitem[{Bar {et~al.}(2019)Bar, Blum, Eby, \& Sato}]{core_halo1}
Bar, N., Blum, K., Eby, J., \& Sato, R. 2019, Phys. Rev. D, 99, 103020

\bibitem[{{Baym} \& {Pethick}(1996)}]{BP}
{Baym}, G., \& {Pethick}, C.~J. 1996, \prl, 76, 6

\bibitem[{Begeman {et~al.}(1991)Begeman, Broeils, \& Sanders}]{PI:1991}
Begeman, K.~G., Broeils, A.~H., \& Sanders, R.~H. 1991, Monthly Notices of the
  Royal Astronomical Society, 249, 523

\bibitem[{Boehmer \& Harko(2007)}]{sf10}
Boehmer, C., \& Harko, T. 2007, Journal of Cosmology and Astroparticle Physics,
  2007, 025

\bibitem[{Broadhurst {et~al.}(2020)Broadhurst, De~Martino, Luu, Smoot, \&
  Tye}]{antlia5}
Broadhurst, T., De~Martino, I., Luu, H.~N., Smoot, G.~F., \& Tye, S.-H.~H.
  2020, Physical Review D, 101, 083012

\bibitem[{Burkert(1995)}]{Burkert:1995}
Burkert, A. 1995, The Astrophysical Journal, 447, L25

\bibitem[{Burkert(2015)}]{scalingmu2}
---. 2015, The Astrophysical Journal, 808, 158

\bibitem[{Burkert(2020)}]{nomu1}
---. 2020, arXiv preprint arXiv:2006.11111

\bibitem[{Calabrese \& Spergel(2016)}]{free_const4}
Calabrese, E., \& Spergel, D.~N. 2016, Monthly Notices of the Royal
  Astronomical Society, 460, 4397

\bibitem[{Caldwell {et~al.}(2017)Caldwell, Walker, Mateo, Olszewski, Koposov,
  Belokurov, Torrealba, Geringer-Sameth, \& Johnson}]{antlia3}
Caldwell, N., Walker, M.~G., Mateo, M., {et~al.} 2017, The Astrophysical
  Journal, 839, 20

\bibitem[{Cede{\~n}o {et~al.}(2020)Cede{\~n}o, Gonz{\'a}lez-Morales, \&
  Ure{\~n}a-L{\'o}pez}]{free_const9}
Cede{\~n}o, F. X.~L., Gonz{\'a}lez-Morales, A.~X., \& Ure{\~n}a-L{\'o}pez,
  L.~A. 2020, arXiv preprint arXiv:2006.05037

\bibitem[{Chan {et~al.}(2018)Chan, Schive, Woo, \& Chiueh}]{chan2018stars}
Chan, J.~H., Schive, H.-Y., Woo, T.-P., \& Chiueh, T. 2018, Monthly Notices of
  the Royal Astronomical Society, 478, 2686

\bibitem[{Chavanis(2011)}]{sc15}
Chavanis, P.-H. 2011, Physical Review D, 84, 043531

\bibitem[{Chavanis(2019)}]{chavanis2019predictive}
---. 2019, Physical Review D, 100, 083022

\bibitem[{Chavanis(2020)}]{chavanis2019core}
---. 2020, Phys. Rev. D, 101, 063532

\bibitem[{Chen {et~al.}(2017)Chen, Schive, \& Chiueh}]{sc17}
Chen, S.-R., Schive, H.-Y., \& Chiueh, T. 2017, Monthly Notices of the Royal
  Astronomical Society, 468, 1338

\bibitem[{Collins {et~al.}(2013)Collins, Chapman, Rich, Ibata, Martin, Irwin,
  Bate, Lewis, Pe{\~n}arrubia, Arimoto, {et~al.}}]{antlia4}
Collins, M.~L., Chapman, S.~C., Rich, R.~M., {et~al.} 2013, The Astrophysical
  Journal, 768, 172

\bibitem[{Davies \& Mocz(2020)}]{davies2020fuzzy}
Davies, E.~Y., \& Mocz, P. 2020, Monthly Notices of the Royal Astronomical
  Society, 492, 5721

\bibitem[{de~Vaucouleurs(1958)}]{vancou1}
de~Vaucouleurs, G. 1958, The Astrophysical Journal, 128, 465

\bibitem[{Di~Paolo {et~al.}(2019)Di~Paolo, Salucci, \& Fontaine}]{salucci}
Di~Paolo, C., Salucci, P., \& Fontaine, J.~P. 2019, The Astrophysical Journal,
  873, 106

\bibitem[{Donato {et~al.}(2009)Donato, Gentile, Salucci, Frigerio~Martins,
  Wilkinson, Gilmore, Grebel, Koch, \& Wyse}]{donato2009constant}
Donato, F., Gentile, G., Salucci, P., {et~al.} 2009, Monthly Notices of the
  Royal Astronomical Society, 397, 1169

\bibitem[{Foreman-Mackey {et~al.}(2013)Foreman-Mackey, Hogg, Lang, \&
  Goodman}]{emcee}
Foreman-Mackey, D., Hogg, D.~W., Lang, D., \& Goodman, J. 2013, Publications of
  the Astronomical Society of the Pacific, 125, 306–312

\bibitem[{{Freeman}(1970)}]{freeman1970ApJ}
{Freeman}, K.~C. 1970, \apj, 160, 811

\bibitem[{Guzm{\'a}n \& Avilez(2018)}]{guzman2018head}
Guzm{\'a}n, F., \& Avilez, A.~A. 2018, Physical Review D, 97, 116003

\bibitem[{Guzm{\'a}n \& Matos(2000)}]{sf14}
Guzm{\'a}n, F.~S., \& Matos, T. 2000, Classical and Quantum Gravity, 17, L9

\bibitem[{Guzm\'an {et~al.}(1999)Guzm\'an, Matos, \& Villegas}]{sf13}
Guzm\'an, F.~S., Matos, T., \& Villegas, H. 1999, Astronomische Nachrichten:
  News in Astronomy and Astrophysics, 320, 97

\bibitem[{Hlozek {et~al.}(2015)Hlozek, Grin, Marsh, \& Ferreira}]{free_const8}
Hlozek, R., Grin, D., Marsh, D.~J., \& Ferreira, P.~G. 2015, Physical Review D,
  91, 103512

\bibitem[{Hu {et~al.}(2000)Hu, Barkana, \& Gruzinov}]{sf2}
Hu, W., Barkana, R., \& Gruzinov, A. 2000, Physical Review Letters, 85, 1158

\bibitem[{Hui {et~al.}(2017)Hui, Ostriker, Tremaine, \& Witten}]{rev4}
Hui, L., Ostriker, J.~P., Tremaine, S., \& Witten, E. 2017, Physical Review D,
  95, 043541

\bibitem[{Ir{\v{s}}i{\v{c}} {et~al.}(2017)Ir{\v{s}}i{\v{c}}, Viel, Haehnelt,
  Bolton, \& Becker}]{free_const7}
Ir{\v{s}}i{\v{c}}, V., Viel, M., Haehnelt, M.~G., Bolton, J.~S., \& Becker,
  G.~D. 2017, Physical review letters, 119, 031302

\bibitem[{Ji \& Sin(1994)}]{sf8}
Ji, S., \& Sin, S.-J. 1994, Physical Review D, 50, 3655

\bibitem[{Lee {et~al.}(2019)Lee, Kim, \& Lee}]{lee2019radial}
Lee, J.-W., Kim, H.-C., \& Lee, J. 2019, Physics Letters B, 795, 206

\bibitem[{Lelli {et~al.}(2017)Lelli, McGaugh, Schombert, \&
  Pawlowski}]{lelli2017one}
Lelli, F., McGaugh, S.~S., Schombert, J.~M., \& Pawlowski, M.~S. 2017, The
  Astrophysical Journal, 836, 152

\bibitem[{Levkov {et~al.}(2018)Levkov, Panin, \& Tkachev}]{sc14}
Levkov, D., Panin, A., \& Tkachev, I. 2018, Physical review letters, 121,
  151301

\bibitem[{Lora \& Magana(2014)}]{lora2}
Lora, V., \& Magana, J. 2014, Journal of Cosmology and Astroparticle Physics,
  2014, 011

\bibitem[{Lora {et~al.}(2012)Lora, Magana, Bernal, S{\'a}nchez-Salcedo, \&
  Grebel}]{lora1}
Lora, V., Magana, J., Bernal, A., S{\'a}nchez-Salcedo, F., \& Grebel, E. 2012,
  Journal of Cosmology and Astroparticle Physics, 2012, 011

\bibitem[{Maga{\~{n}}a \& Matos(2012)}]{rev1}
Maga{\~{n}}a, J., \& Matos, T. 2012, Journal of Physics: Conference Series,
  378, 012

\bibitem[{Maleki {et~al.}(2020)Maleki, Baghram, \&
  Rahvar}]{maleki2020constraint}
Maleki, A., Baghram, S., \& Rahvar, S. 2020, Physical Review D, 101, 103504

\bibitem[{Marsh(2016)}]{rev3}
Marsh, D.~J. 2016, Physics Reports, 643, 1

\bibitem[{Marsh \& Ferreira(2010)}]{sf11}
Marsh, D.~J., \& Ferreira, P.~G. 2010, Physical Review D, 82, 103528

\bibitem[{Marsh \& Pop(2015)}]{sc16}
Marsh, D.~J., \& Pop, A.-R. 2015, Monthly Notices of the Royal Astronomical
  Society, 451, 2479

\bibitem[{Matos {et~al.}(2000)Matos, Guzm{\'a}n, \& Ure\~na L{\'o}pez}]{sf4}
Matos, T., Guzm{\'a}n, F.~S., \& Ure\~na L{\'o}pez, L.~A. 2000, Classical and
  Quantum Gravity, 17, 1707

\bibitem[{Matos \& Ure\~na L\'opez(2000)}]{sf40}
Matos, T., \& Ure\~na L\'opez, L.~A. 2000, Classical and Quantum Gravity, 17,
  L75

\bibitem[{Matos \& Ure\~na L\'opez(2001)}]{sf3}
---. 2001, Physical Review D, 63, 063506

\bibitem[{{McGaugh} {et~al.}(2016){McGaugh}, {Lelli}, \&
  {Schombert}}]{McGaugh:2016PRL}
{McGaugh}, S.~S., {Lelli}, F., \& {Schombert}, J.~M. 2016, Physical Review
  Letters, 117, 201101

\bibitem[{Membrado {et~al.}(1989)Membrado, Pacheco, \& Sa{\~n}udo}]{sf12}
Membrado, M., Pacheco, A., \& Sa{\~n}udo, J. 1989, Physical Review A, 39, 4207

\bibitem[{Mina {et~al.}(2020)Mina, Mota, \& Winther}]{nomu2}
Mina, M., Mota, D.~F., \& Winther, H.~A. 2020, arXiv preprint arXiv:2007.04119

\bibitem[{Mocz {et~al.}(2017)Mocz, Vogelsberger, Robles, Zavala,
  Boylan-Kolchin, Fialkov, \& Hernquist}]{sc13}
Mocz, P., Vogelsberger, M., Robles, V.~H., {et~al.} 2017, Monthly Notices of
  the Royal Astronomical Society, 471, 4559

\bibitem[{Newville {et~al.}(2014)Newville, Stensitzki, Allen, \&
  Ingargiola}]{lmfit}
Newville, M., Stensitzki, T., Allen, D.~B., \& Ingargiola, A. 2014, {LMFIT:
  Non-Linear Least-Square Minimization and Curve-Fitting for Python}, v.0.8.0,
  Zenodo, doi:10.5281/zenodo.11813

\bibitem[{Niemeyer(2020)}]{niemeyer2019small}
Niemeyer, J.~C. 2020, Progress in Particle and Nuclear Physics, 113, 103787

\bibitem[{Padilla {et~al.}(2020)Padilla, Rindler-Daller, Shapiro, Matos, \&
  V{\'a}zquez}]{mipaper}
Padilla, L.~E., Rindler-Daller, T., Shapiro, P.~R., Matos, T., \& V{\'a}zquez,
  J.~A. 2020, arXiv preprint arXiv:2010.12716

\bibitem[{Padilla {et~al.}(2019)Padilla, Tellez, Escamilla, \&
  Vazquez}]{padilla2019cosmological}
Padilla, L.~E., Tellez, L.~O., Escamilla, L.~A., \& Vazquez, J.~A. 2019, arXiv
  preprint arXiv:1903.11127

\bibitem[{Rindler-Daller \& Shapiro(2014)}]{RS}
Rindler-Daller, T., \& Shapiro, P.~R. 2014, Modern Physics Letters A, 29,
  1430002

\bibitem[{Robles \& Matos(2012)}]{robles2012exact}
Robles, V.~H., \& Matos, T. 2012, The Astrophysical Journal, 763, 19

\bibitem[{Ruffini \& Bonazzola(1969)}]{wf1}
Ruffini, R., \& Bonazzola, S. 1969, Physical Review, 187, 1767

\bibitem[{Sahni \& Wang(2000)}]{sf1}
Sahni, V., \& Wang, L. 2000, Physical Review D, 62, 103517

\bibitem[{Salucci \& Burkert(2000)}]{scalingmu1}
Salucci, P., \& Burkert, A. 2000, The Astrophysical Journal Letters, 537, L9

\bibitem[{Sarkar {et~al.}(2016)Sarkar, Mondal, Das, Sethi, Bharadwaj, \&
  Marsh}]{free_const5}
Sarkar, A., Mondal, R., Das, S., {et~al.} 2016, Journal of Cosmology and
  Astroparticle Physics, 2016, 012

\bibitem[{Schive {et~al.}(2014{\natexlab{a}})Schive, Chiueh, \&
  Broadhurst}]{sc4}
Schive, H.-Y., Chiueh, T., \& Broadhurst, T. 2014{\natexlab{a}}, Nature
  Physics, 10, 496

\bibitem[{Schive {et~al.}(2014{\natexlab{b}})Schive, Liao, Woo, Wong, Chiueh,
  Broadhurst, \& Hwang}]{sc10}
Schive, H.-Y., Liao, M.-H., Woo, T.-P., {et~al.} 2014{\natexlab{b}}, Physical
  review letters, 113, 261302

\bibitem[{Schwabe {et~al.}(2016)Schwabe, Niemeyer, \& Engels}]{sc11}
Schwabe, B., Niemeyer, J.~C., \& Engels, J.~F. 2016, Physical Review D, 94,
  043513

\bibitem[{Sin(1994)}]{sf80}
Sin, S.-J. 1994, Physical Review D, 50, 3650

\bibitem[{Sofue(2013)}]{MWdata}
Sofue, Y. 2013, Publications of the Astronomical Society of Japan, 65,
  https://academic.oup.com/pasj/article-pdf/65/6/118/6030872/pasj65-0118.pdf,
  118

\bibitem[{Sofue(2017)}]{effects1}
---. 2017, Publications of the Astronomical Society of Japan, 69, R1

\bibitem[{Sofue {et~al.}(2009)Sofue, Honma, \& Omodaka}]{vancou2}
Sofue, Y., Honma, M., \& Omodaka, T. 2009, Publications of the Astronomical
  Society of Japan, 61, 227

\bibitem[{Spano {et~al.}(2008)Spano, Marcelin, Amram, Carignan, Epinat, \&
  Hernandez}]{spano2008ghasp}
Spano, M., Marcelin, M., Amram, P., {et~al.} 2008, Monthly Notices of the Royal
  Astronomical Society, 383, 297

\bibitem[{Strigari {et~al.}(2008)Strigari, Bullock, Kaplinghat, Simon, Geha,
  Willman, \& Walker}]{NatureBullock}
Strigari, L.~E., Bullock, J.~S., Kaplinghat, M., {et~al.} 2008, Nature, 454,
  1096

\bibitem[{Su{\'a}rez {et~al.}(2014)Su{\'a}rez, Robles, \& Matos}]{rev2}
Su{\'a}rez, A., Robles, V.~H., \& Matos, T. 2014, in Accelerated Cosmic
  Expansion (Springer), 107--142

\bibitem[{Torrealba {et~al.}(2016)Torrealba, Koposov, Belokurov, \&
  Irwin}]{antlia2}
Torrealba, G., Koposov, S., Belokurov, V., \& Irwin, M. 2016, Monthly Notices
  of the Royal Astronomical Society, 459, 2370

\bibitem[{Torrealba {et~al.}(2019)Torrealba, Belokurov, Koposov, Li, Walker,
  Sanders, Geringer-Sameth, Zucker, Kuehn, Evans, {et~al.}}]{antlia1}
Torrealba, G., Belokurov, V., Koposov, S., {et~al.} 2019, Monthly Notices of
  the Royal Astronomical Society, 488, 2743

\bibitem[{Ure\~na L\'opez {et~al.}(2017)Ure\~na L\'opez, Robles, \&
  Matos}]{Urena-Robles-Matos:2017}
Ure\~na L\'opez, L.~A., Robles, V.~H., \& Matos, T. 2017, Phys. Rev. D, 96,
  043005

\bibitem[{Veltmaat \& Niemeyer(2016)}]{sc12}
Veltmaat, J., \& Niemeyer, J.~C. 2016, Physical Review D, 94, 123523

\bibitem[{Veltmaat {et~al.}(2020)Veltmaat, Schwabe, \&
  Niemeyer}]{veltmaat2020baryon}
Veltmaat, J., Schwabe, B., \& Niemeyer, J.~C. 2020, Physical Review D, 101,
  083518

\end{thebibliography}
\bibliographystyle{aasjournal}

\end{document}